\documentclass[12pt]{iopart}
\usepackage{epsfig}
\begin{document}
\topical[Critical phenomena at perfect and non-perfect surfaces]{Critical phenomena
at perfect and non-perfect surfaces}

\author{M Pleimling}
\address{Institut f\"ur Theoretische Physik I,
Universit\"at Erlangen-N\"urnberg, D -- 91058 Erlangen, Germany}

\ead{pleim@theorie1.physik.uni-erlangen.de}

\begin{abstract}
In the past perfect surfaces have been shown to yield a local critical
behaviour that differs from the bulk critical behaviour.
On the other hand surface defects, whether they are of natural
origin or created artificially, are known to modify local quantities. It is
therefore important to clarify whether these defects are relevant or
irrelevant for the surface critical behaviour.

The purpose of this review is two-fold. In the first part we summarise some
of the important results on surface criticality at perfect surfaces. Special
attention is thereby paid to new developments as for example the study of surface critical
behaviour in systems with competing interactions or of surface critical dynamics.
In the second part the effect of surface defects (presence of edges, steps,
quenched randomness, lines of adatoms, regular geometric patterns) on local critical behaviour
in semi-infinite systems and in thin films
is discussed in detail. Whereas most of the defects commonly encountered are 
shown to be irrelevant, some notable exceptions are highlighted. It is shown furthermore that under
certain circumstances non-universal local critical behaviour may be observed at surfaces.
\end{abstract}

\submitto{\JPA}

\maketitle

\section{Introduction}
Our current understanding of critical phenomena results from an intensive interplay
between experimental studies of a large variety of physical systems, ground-breaking
theoretical developments (including renormalization group methods, finite-size
scaling theory and conformal invariance) and extensive numerical investigations
of model systems. In many cases the systems under investigation are treated
as bulk systems, thus neglecting the existence of surfaces which are unavoidable in real
physical systems.
Discarding surfaces in systems with short-range interactions is justifiable when studying 
bulk critical properties,
as the contribution of the surface to extensive quantities is
vanishing in the thermodynamic limit. However, a surface breaks the translation
symmetry of a system and changes local quantities.
Thirty years ago, it has been realized that this leads to surface critical
behaviour which differs
from the bulk critical behaviour. Since that time numerous theoretical
and experimental studies have been undertaken in order to determine local critical quantities
in systems with perfect surfaces.

Real surfaces, however,
are usually not perfectly smooth but display some degree of roughness
due to the presence of surface defects as for example steps, islands,
or vacancies. Impurities are also often encountered at crystalline surfaces
and may be viewed as the source of some disorder at the surface. Furthermore, experimentalists
nowadays create thin films which do not appear in nature, by growing artificial
structures on the film surface. All these defects have some impact on
magnetic surface quantities.

The present work reviews the recent progress achieved in the study of critical phenomena 
in systems with boundaries. Besides semi-infinite systems
and films with perfect surfaces, more complex geometries with wedges and corners as well as more
realistic surfaces with defects are discussed. The review hereby focuses
on the question whether the different types of geometries and/or of surface defects have an impact
on the surface critical behaviour. It is therefore complementary to earlier reviews
on surface criticality \cite{Bin83,Abr86,Die86,Die97} that exclusively considered flat, perfect surfaces.

The thermodynamics
of a surface is completely described by the surface free energy per area $F_s$. Singularities
occurring in $F_s$ determine the phase diagram of semi-infinite systems.
At some phase boundaries of this phase diagram
surface and bulk free energies both exhibit singularities, 
whereas at other boundaries only $F_s$ becomes singular.
An example for the former case is the {\bf ordinary transition} where bulk and 
surface ordering occur
at the same temperature, whereas the latter case is encountered at the so-called {\bf surface transition}
where the surface layer alone orders, while the bulk remains disordered. 
This surface transition is encountered at temperatures higher than the bulk critical
temperature, the critical fluctuations of the $d$-dimensional semi-infinite system
then being essentially ($d-1$)-dimensional, corresponding to a phase transition
in $d-1$ dimensions.

In Section 2
the critical behaviour at perfect surfaces is discussed.
At the bulk critical point
different surface universality classes are obtained for every bulk universality class. 
These universality classes 
are discussed and their
differences emphasized. 
Furthermore, recent progress in our understanding of surface critical behaviour
in systems with competing interactions is reviewed. Surface critical dynamics and the
effect of symmetry breaking surface fields are also briefly discussed.
Section 3 is devoted to more complex geometries with a wedge.
Wedge-shaped models
are very interesting systems where
local critical exponents, which change continuously with the wedge
opening angle, arise because
of the particular geometrical properties of the wedge.
However, at a given opening angle,
surface critical behaviour at the ordinary transition is still universal and does not 
depend on microscopic
details of the
model as for example the lattice type or the strengths of the local interactions. 
This is completely
different at the surface transition, where under certain circumstances
non-universal local critical behaviour 
can be observed. At a fixed opening angle, edge critical exponents
then not only depend continuously on the
values of the local couplings but also reflect the existence of the disordered bulk. 
This intriguing
behaviour results from the fact that at the surface transition the edge acts like a defect line
in a two-dimensional critical system. Corner critical behaviour is discussed in this 
Section, too. 
In the last years a great deal of activity focused on the critical behaviour of wedges in
presence of external surface fields. This rapidly developing field is also briefly
summarized.

Section 4 deals with the important issue of thin films and semi-infinite 
systems with non-perfect
surfaces. 
Surfaces are very often naturally rough, due to the growth mechanism or because of
erosion effects. Adatom islands, vacancy islands or steps are typical defects encountered
at real surfaces. On the other hand, specific surface structures, as for example lines
of adatoms or regular geometrical patterns, can be created on purpose by using advanced 
experimental methods.
As these surface defects have
an impact on local surface quantities, one has to ask the question whether they do change
the surface critical behaviour. For semi-infinite systems we must again distinguish between
the ordinary transition and the surface transition. At the ordinary transition, 
common surface defects
(presence of a step, surfaces with uncorrelated roughness, amorphous surface) are
usually irrelevant for the surface critical behaviour,
but there are some notable exceptions. At the surface transition, 
defect structures
like steps and additional lines of atoms located at the surface may yield non-universal local
critical exponents, as observed in semi-infinite Ising models with additional surface structures.
Interestingly, additional
lines located at the surface of thin Ising films always lead to a 
non-universal local critical behaviour. Section 5 finally contains
concluding remarks.

\section{Perfect surfaces}
The aim of this Section is to review the main results on surface criticality in systems
with flat, perfect surfaces.
In Sections 2.1 and 2.2
we pay special attention
to the phase diagrams of semi-infinite systems as well as to the sets of
critical exponents characterizing the different surface phase transitions.
As the amount of papers published
on this topic is rather impressive, only selected results are presented. The interested reader is
referred to the reviews of Binder \cite{Bin83}, Abraham \cite{Abr86}, and Diehl \cite{Die86, Die97}
for further readings. The main experimental techniques available
for the investigation of surface critical behaviour are reviewed in the book of Dosch \cite{Dos92}. 
Section 2.3 is devoted to critical phenomena in thin films, whereas in Section 2.4 the recent
studies of surface criticality in systems with competing interactions are discussed. Brief
accounts of works on surface critical dynamics and on surface critical behaviour in presence
of external fields close this Section.

\subsection{Surface quantities and phase diagrams}
When comparing systems with and without surfaces, one remarks that extensive quantities 
are altered by the
presence of surfaces. Thus, the free energy is given by \cite{Fis77, Cag79}
\begin{equation} \label{gl:3_2}
F = F_b \, V + F_s \, S + \ldots
\end{equation}
with the bulk free energy per volume $F_b$ and the surface free energy per 
surface area $F_s$. $V$ and $S$
are the volume and the surface of the system, respectively. Further terms in 
equation (\ref{gl:3_2}) may result
from the presence of edges, as discussed in Section 3. 
Similar corrections also show up in other extensive quantities. 

The surface free energy $F_s$ completely describes the thermodynamics of the surface.
Similar to $F_b$, singularities occur in $F_s$. The values of the variables of $F_s$, 
where these singularities appear, determine the boundaries 
separating the different phases of the surface phase diagram. As we shall see later in the course
of this Section, at some phase boundaries
both $F_b$ and $F_s$ exhibit singularities whereas at other ones only $F_s$ becomes singular.

\begin{figure} 
\centerline{
\psfig{figure=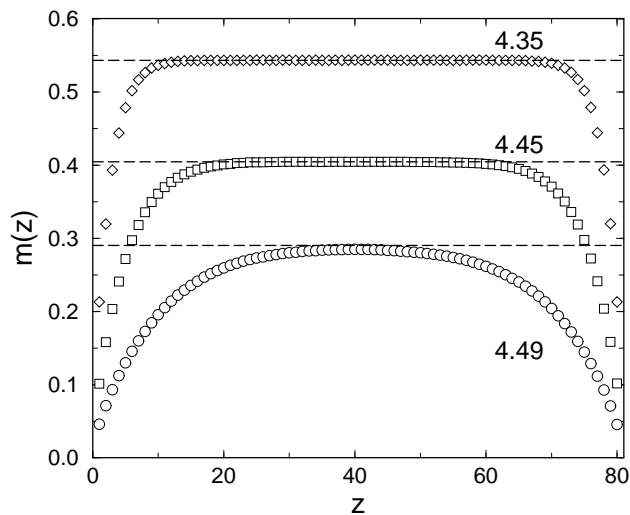,width=7cm,angle=270}
}
\caption
{\label{fig_surf_1} Order parameter profiles $m(z)$ of a three-dimensional Ising film with 80 layers
at three different temperatures $k_BT/J_b$.
The dashed lines denote the bulk values. After \cite{Ple98a}.
}
\end{figure}

Local quantities as the local magnetisation density (often called local magnetisation for brevity)
or the local energy density are also modified by the presence of a surface.
Examples of order parameter profiles
obtained by Monte Carlo simulations of three-dimensional 
Ising films with $L=80$ layers at temperatures below the
bulk critical temperature $k_BT_{c}/J_b=4.5115$
are shown in Figure \ref{fig_surf_1} \cite{Ple98a}. Here, all the couplings involved
have been chosen to have equal strength $J_b$ and no magnetic fields have been retained.
The surface magnetisation $m_1=m(1)=m(L)$, related to $F_s$ by the equation
\begin{equation} \label{gl:3_3}
m_1= - \frac{\partial F_s}{\partial H_1},
\end{equation}
is smaller than the bulk magnetisation due to the reduced coordination number at the surface.
The field $H_1$ acts exclusively on the surface layer.
The magnetisation increases from its surface value to the bulk value at distances exceeding the bulk
correlation length $\xi_b$. In case the thickness of the film is not large 
compared to $\xi_b$ (as for example
at the temperature $k_BT/J_b=4.49$ in Figure \ref{fig_surf_1})
the bulk magnetisation is never reached. For surface couplings
exceeding the bulk couplings by a large amount, magnetisation profiles decreasing monotonically towards
the center of the system may also be observed. 

For Ising models in film geometry
we have the general Hamiltonian
\begin{equation} \label{gl:3_1}
{\mathcal H}=- J_b \sum\limits_{\left< i \, j \right>} \, s_i s_j \, \, - \, J_s \sum\limits_{
{\scriptsize
\begin{array}{c}
\left< i \, j \right>\\surface
\end{array}}}
\, s_i s_j \, \, - H \sum\limits_i \, s_i \, \, - H_1 \sum\limits_{surface} s_i.
\end{equation}
where the spins $s_i$ can take on the values $\pm 1$.
The first sum runs over bonds connecting neighbouring spins where at least one of the spins 
is a bulk spin,
whereas the second sum runs over all surface links. Bulk couplings,
$J_b$, and surface couplings, $J_s$, are ferromagnetic. The bulk field $H$ acts on all the spins, 
the surface
field $H_1$ only on spins located at the surface, i.e. a surface spin sees the field
$H + H_1$. Sometimes a different coupling constant is used for the
bonds connecting surface and bulk spins \cite{Lub75}. In general, 
the perturbations due to the presence of
a surface are supposed to be of short range.

Analytical results on surface critical behaviour are commonly obtained in the framework
of continuum field theory, see \cite{Die86,Die97} for comprehensive reviews. The standard
$\phi^4$ Ginzburg-Landau model, appropriate for the universality class of the Ising model,
is thereby augmented in order to include contributions coming from the surface:
\begin{equation} \label{gl:field}
f_s := F_s/k_BT = \frac{1}{2} c_0 \, \phi^2 - h_1 \, \phi 
\end{equation}
with $h_1:=H_1/k_BT$,
whereas $c_0$ is related to the surface enhancement of the spin-spin coupling constant
in the corresponding lattice model. The resulting Ginzburg-Landau free energy density 
is readily generalized to cases where the order parameter exhibits a different symmetry.

Layer-dependent quantities are very useful when investigating systems with surfaces.
Examples are the magnetisation per layer $m(z)$ 
and the susceptibility per layer $\chi(z)$ where $z$ labels the layers parallel to the surface.
The surface magnetisation is
$m_1=m(z=1)$ and the local susceptibility at the surface, i.e.\ 
the response of the surface magnetisation to
a surface field, $\chi_{11}= - \frac{\partial^2 F_s}{\partial H_1^2}$,
is given by $\chi_{11}=\chi(z=1)$. From the profiles $m(z)$
and $\chi(z)$ one also obtains the surface excess quantities
\begin{equation} \label{gl:3_3c}
m_s=- \frac{\partial F_s}{\partial H}=\sum\limits_{z=1} \left(m(z) - m_b \right)
\end{equation}
and
\begin{equation} \label{gl:3_3d}
\chi_s=- \frac{\partial^2 F_s}{\partial H^2}=
\sum\limits_{z=1} \left( \chi(z) - \chi_b \right)
\end{equation}
where $m_b$ and $\chi_b$ are the bulk magnetisation and the bulk susceptibility.
Of further importance is the surface layer susceptibility, i.e.\
the response of the local surface magnetisation to a bulk field, usually
denoted by $\chi_1$. 
The critical surface pair correlation function behaves as
\begin{equation} \label{gl:gpar}
G_{par}(\vec{\rho}-\vec{\rho}\,') \sim \left| \vec{\rho}-\vec{\rho}\,' \right|^{-2 x_1}
\end{equation}
where $x_1$ is the scaling dimension of the surface order parameter.
For a $d$-dimensional model $\vec{\rho}$
is a $(d-1)$-dimensional vector parallel to the surface. This behaviour of the
surface correlation function differs from that of the bulk correlation function:
\begin{equation} \label{gl:gbulk}
G_b(\vec{r} -\vec{r}\,') \sim \left| \vec{r} -\vec{r}\,' \right|^{-2 x_b}
\end{equation}
where the value of the bulk scaling dimension $x_b$ is usually smaller than that of $x_1$. 
Here $\vec{r}$ is a $d$-dimensional vector.
Critical correlations between a surface point and a bulk point exhibit a power law behaviour
with the exponent $x_1+x_b$.
The surface
critical exponents associated with the different critical
quantities are listed in Table \ref{table_surf_1}.

\begin{table}
\begin{center}
\begin{tabular}{|c|c|}
\hline
Surface exponent & Definition \\
\hline\hline
$\beta_1$ & $m_1 \sim t^{\beta_1}$ \\
\hline
$\gamma_1$ & $\chi_1 \sim \left| t \right|^{-\gamma_1}$ \\
\hline
$\gamma_{11}$ & $\chi_{11} \sim \left| t \right|^{-\gamma_{11}}$ \\
\hline
$\beta_s$ & $m_s \sim t^{\beta_s}$ \\
\hline
$\gamma_s$ & $\chi_s \sim \left| t \right|^{-\gamma_s}$ \\
\hline
$\alpha_s$ & $C_s  \sim \left| t \right|^{-\alpha_s}$ \\
\hline
$x_1$ & $G_{par}(\vec{\rho}-\vec{\rho}\,') \sim \left| \vec{\rho}-\vec{\rho}\,' \right|^{-2 x_1}$ \\
\hline
$\Phi$ & $\left| T_s(c)-T_{c} \right|/T_{c} \sim \left| c \right|^{1/\Phi}$ \\
\hline
\end{tabular}
\caption
{\label{table_surf_1} Surface critical exponents and their definitions.}
\end{center}
\end{table}

\begin{figure} 
\centerline{
\psfig{figure=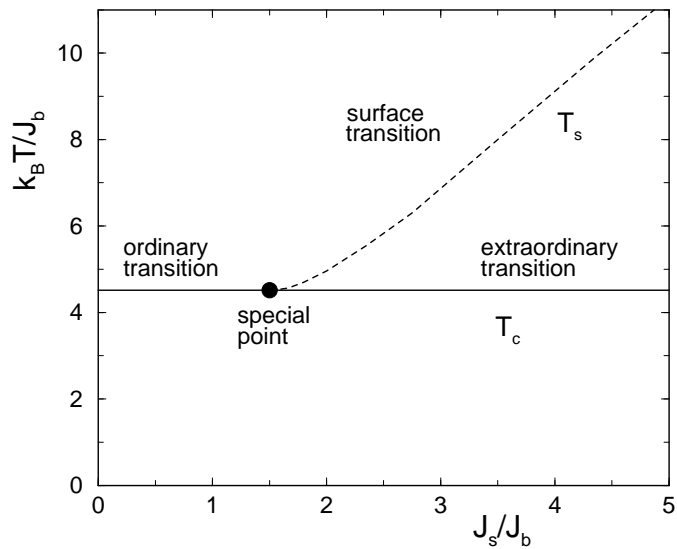,width=7cm,angle=270}
}
\caption
{\label{fig_surf_2} Surface phase diagram of the semi-infinite three-dimensional Ising model.
}
\end{figure}

The phase diagram of the three-dimensional semi-infinite Ising model is well established.
The global phase diagram of the semi-infinite system depends not only 
on the values of the coupling constants and on the
temperature but also on the external fields $H$ and $H_1$ \cite{Nak82}.
I discuss in the following only the case of vanishing external fields.
The qualitative results obtained in mean-field approximation \cite{Mil71, Bin72, Bin74, Lub75, Bin83}
represent the correct
phase diagram as derived from renormalization group calculations \cite{Bur77, Lip81}
and from Monte Carlo
simulations \cite{Bin74, Bin84} quite well, see Figure \ref{fig_surf_2}.
If the ratio of the surface coupling $J_s$ to the bulk coupling $J_b$, $r=J_s/J_b$, 
is sufficiently small, the system undergoes at the bulk critical temperature
$T_{c}$ an ordinary transition, with the bulk and surface ordering occurring
at the same temperature.
Beyond a critical ratio, $r > r_{sp} \approx 1.50$ for the semi-infinite Ising model
on the simple cubic lattice \cite{Bin84,Rug92,Rug93},
the surface orders at the so-called surface transition at a
temperature $T_s > T_{c}$, followed by the extraordinary transition of
the bulk at $T_{c}$. At the critical ratio $r_{sp}$, one encounters the multicritical
special transition point,
with critical surface properties deviating from those at the ordinary transition
and those at the surface transition. In mean-field approximation
the critical ratio at the special transition point is $r_{sp}^{MF}=1.25$ for the simple
cubic lattice.

In the field theoretical formulation of the problem, see Eq.\ (\ref{gl:field}), three stable
renormalisation-group fixed-point values for $c_0$ can be shown to exist in absence 
of external fields. For $c_0  = + \infty$ resp.\ $c_0 = - \infty$ one is dealing with the
critical behaviour of the ordinary resp.\ extraordinary transition. The fixed point corresponding
to the special transition is located at $c_0 = c_{SP}$. Usually, a new variable $c \sim
c_0 - c_{SP}$ is introduced, such that the ordinary transition occurs for $c>0$, whereas
the extraordinary transition is observed for supercritical enhancements $c < 0$, the 
special transition point then being located at $c=0$.

Mean-field approximation yields similar surface phase diagrams for all $O(n)$ models in all dimensions.
Of course, this scenario with three lines of continuous phase transitions is only realized in
systems where the surface can sustain long-range order, independently of the bulk. Rigorous
results \cite{AuY73} show that
only the ordinary transition is encountered in
the two-dimensional Ising model with short-range interactions
in absence of external fields.
This is readily understood as the surface is then one-dimensional.
For continuous spins with $n \geq 3$ and isotropic couplings,
a surface transition, where only the surface orders, does not
exist in three dimensions, in accordance with the well known Mermin-Wagner theorem \cite{Mer66}.
Indeed, the surface effectively decouples from the bulk for very 
strong surface couplings $J_s \gg J_b$
and then forms an isolated two-dimensional system. An interesting case is that of 
the $XY$ ($n=2$) model
which has a Kosterlitz-Thouless transition in two dimensions \cite{Kos73, Kos74}.
A phase diagram similar
to Figure \ref{fig_surf_2} has been shown to exist for the three-dimensional
semi-infinite $XY$ model \cite{Fro86, Lan89,Pec91}
with the surface undergoing a Kosterlitz-Thouless transition for coupling ratios 
larger then some critical value,
the bulk remaining disordered. Due to the peculiar properties of this transition,
the multicritical special transition point is then of a special nature. Diehl and Eisenriegler
have shown that in semi-infinite $O(n)$ models (and therefore also in the experimental
relevant case
of the semi-infinite Heisenberg model in three dimensions) anisotropic surface
couplings may lead to
an anisotropic special transition point and to a surface transition where the
bulk remains disordered \cite{Die84}. Phase diagrams similar to Figure \ref{fig_surf_2}
are found for all $O(n)$ models
with isotropic couplings above the upper critical dimension.

The models mentioned so far only cover a fraction of the research on surface critical phenomena.
Other semi-infinite models studied include, for example, spin-1 Ising models \cite{Aff00},
Blume-Emery-Griffiths models \cite{Bak01}, Potts models \cite{Igl99},
ferrimagnetic Ising models \cite{Ben97}, layered magnetic systems \cite{Kar97} or
bond percolation in the semi-infinite system (which can be regarded as the
($q \longrightarrow 1$)-limit of the $q$-state Potts model) \cite{The79,DeB80,
Car80,Chr86,Die89}.

Before discussing the various surface universality classes, I want to mention two 
further interesting cases.
For systems with a discontinuous phase transition, the surface order parameter may 
change continuously as the bulk transition point is approached.
As discussed by Lipowsky \cite{Lip82} (see also \cite{Laj81})
universal surface properties may show up for this case too. In the presence of this so-called
surface-induced disorder, correlation lengths both parallel and perpendicular to the surface
diverge at the bulk first-order transition point, thus inducing anisotropic power law
behaviour for some bulk quantities \cite{Tur02}. A possible realization of this
scenario may be encountered in the antiferromagnet UO$_2$ where the surface layers
order continuously while the bulk displays a discontinuous ordering \cite{Wat00}.
A second intriguing scenario is encountered in the $q$-state Potts model \cite{Lip82b,Dob04}
where a phase transition to a low temperature phase with an ordered bulk but a 
disordered surface is observed.

\subsection{The surface universality classes}
It is obvious from the preceding discussion of possible surface phase diagrams
that for every bulk universality
class different surface universality classes may be realized. 
The different surface universality
classes will be discussed in the following in more detail and their differences emphasized.

At the ordinary transition both bulk and surface order at the bulk critical temperature.
The ordinary transition is the only phase transition which is observed in the surface phase diagrams
of all ferromagnetic three-dimensional $O(n)$ models in absence of
symmetry-breaking fields. It has been studied intensively, both by
analytical and by numerical methods. The surface critical exponents at the ordinary transition
can all be obtained by combining bulk exponents with one additional surface exponent $\Delta_1$
\cite{Bin72,Bin74}.
This new exponent results from the scaling function of the singular part of the
surface free energy
\begin{equation} \label{gl:3_3e}
f_s^{(sing)}= \left| t \right|^{2 - \alpha_s} \, g\left(  \left| t \right|^{-\Delta_b} h,
\left| t \right|^{-\Delta_1} \tilde{h}_1 \right)
\end{equation}
where $t=(T_{c}-T)/T_{c}$ is the reduced temperature and
$\tilde{h}_1$ is the surface scaling field which depends on
both bulk and surface fields. The bulk exponent $\Delta_b$ is known from the
singular part of the bulk free energy, whereas $\alpha_s$ is the critical exponent
of the excess specific heat $C_s$. Various scaling relations connect surface and bulk
exponents. The critical exponents of excess quantities are obtained by
combining bulk exponents:
\begin{equation} \label{gl:3_3f}
\beta_s=\beta_b - \nu_b
\end{equation}
\begin{equation} \label{gl:3_3g}
\gamma_s=\gamma_b + \nu_b
\end{equation}
\begin{equation} \label{gl:3_3h}
\alpha_s= \alpha_b - \nu_b
\end{equation}
Other useful scaling relations are ($d$ being the number of space dimensions)
\begin{equation} \label{gl:3_3i}
\gamma_s= 2 \gamma_1 - \gamma_{11}
\end{equation}
\begin{equation} \label{gl:3_3j}
\gamma_{11}= \nu_b \left(d -1 - 2 x_1  \right)
\end{equation}
\begin{equation} \label{gl:3_3k}
\gamma_1 = \nu_b \left(d - x_1 -x_b \right)
\end{equation}
\begin{equation} \label{gl:3_3l}
\beta_1 = \nu_b  \, x_1
\end{equation}
Further scaling relations are discussed in \cite{Bin74,Bin83,Die86}.

In Table \ref{table_surf_2} estimates of various three-dimensional Ising surface critical exponents
obtained by different techniques are compiled. One observes a very good agreement
between the recent massive field-theoretical estimates \cite{Die94,Die98a}
and the numerical estimates. Similar agreement is also obtained for other
$O(n)$ models. From the numerical point of view, the best investigated cases are
$n=0$ \cite{DeB90,Heg94} and $n=1$ (Ising) \cite{Lan90,Rug92,Rug93,Rug95,Ple98a},
whereas studies for $n \geq 2$ are scarce \cite{Lan89,Nig88,Kre00,Ber03}.
The rather few experimental determinations of surface critical exponents
at the ordinary transition, using for example x-ray scattering at grazing angle,
yield values which are found to be compatible with the theoretical estimates 
\cite{Sig86,Mai90,Bur93,Alv82}. 

\begin{table}
\begin{center}
\begin{tabular}{|c||c|c|c|}
\hline
 & $\beta_1$ & $\gamma_1$ & $\gamma_{11}$\\
\hline\hline
MF \cite{Bin72} & 1 & 1/2 & $-$1/2 \\
\hline
MC \cite{Lan90} & $0.78(2)$ & $0.78(6)$ & \\
\hline
MC \cite{Rug95} & $0.807(4)$ & $0.760(4)$ & \\
\hline
MC \cite{Ple98a} & $0.80(1)$ & $0.78(5)$ & $-0.25(10)$ \\
\hline
FT \cite{Die81a} & 0.816 & 0.767 & $-0.336$ \\
\hline
FT \cite{Die98a} & 0.796 & 0.769 & $-0.333$ \\
\hline
\end{tabular}
\caption
{\label{table_surf_2} 
Estimates of Ising surface critical exponents in three dimensions
at the ordinary transition, as obtained by
different techniques. MF: mean-field theory, MC: Monte Carlo techniques, FT: field-theoretical
methods.}
\end{center}
\end{table}

Exact results can by obtained by applying conformal invariance to semi-infinite systems.
As shown by Cardy \cite{Car84a} conformal invariance yields in two-dimensional
systems the exact values of the surface critical exponents at the ordinary transition
and completely fixes the correlation functions. In higher dimensions it constrains
the form of correlation functions near the free surface. Recent applications of conformal
invariance include the determination of order parameter profiles in various
two-dimensional systems with boundaries \cite{Res00} or the computation of 
three-dimensional Ising surface critical exponents from models defined on
half spherocylinders \cite{Den03}.

The special transition point, located at the coupling ratio
$r_{sp}$, is a multicritical point where the bulk {\it and}
the surface become critical.
Three different critical lines
(ordinary transition, surface transition, extraordinary transition) merge at this point.
The surface criticality is thereby characterized by two new surface exponents:
$\Delta_1^{sp}$ and $\Phi$. The appropriate scaling ansatz for the singular part
of the surface free energy then reads:
\begin{equation} \label{gl:3_3n}
f_s^{(sing)} = \left| t \right|^{2 - \alpha_s} \, g_{sp} \left(  \left| t \right|^{-\Delta_b} h,
\left| t \right|^{-\Delta_1^{sp}} h_1, \left| t \right|^{- \Phi} \, c  \right)
\end{equation}
with the surface enhancement $c \sim r - r_{sp}$.
The crossover exponent $\Phi$ governs the behaviour of the surface transition line close
to the special transition point:
\begin{equation} \label{gl:3_3o}
\left| T_s(c) - T_{c} \right|/T_{c} \sim \left| c \right|^{1/\Phi}.
\end{equation}
Estimates for various surface critical exponents obtained at the special transition point
of the three-dimensional semi-infinite Ising model are gathered in Table \ref{table_surf_3}. 
Again, good agreement between analytical and numerical estimates
has been achieved. One also notes that the scaling laws (\ref{gl:3_3j})-(\ref{gl:3_3l})
hold at this multicritical point.

\begin{table}
\begin{center}
\begin{tabular}{|c||c|c|c|c|}
\hline
 & $\beta_1^{sp}$ & $\gamma_1^{sp}$ & $\gamma_{11}^{sp}$ & $\Phi$\\
\hline\hline
MF \cite{Bin83} & 1/2 & 1 & 1/2 & 1/2 \\
\hline
MC \cite{Lan90} & $0.18(2)$ & $1.41(14)$ & $0.96(9)$ & $0.59(4)$ \\
\hline
MC \cite{Rug93} & $0.237(5)$ & & & $0.461(15)$ \\
\hline
MC \cite{Rug95} & $0.2375(15)$ & $1.328(1)$ & $0.788(1)$ & \\
\hline
FT \cite{Die81b} & 0.245 & 1.43 & 0.85 & 0.68\\
\hline
FT \cite{Die98a} & 0.263 & 1.302 & 0.734 & 0.539 \\
\hline
\end{tabular}
\caption
{\label{table_surf_3}
Estimates of Ising surface critical exponents at the special transition point in three
dimensions, as obtained by
different techniques. MF: mean-field theory, MC: Monte Carlo techniques, FT: field-theoretical
methods.}
\end{center}
\end{table}

For surface enhancements exceeding the critical value $r_{sp}$ two distinct phase transitions
are encountered. The surface transition line, which is located at higher temperatures,
separates the disordered high temperature phase from a phase where the surface alone is ordered.
The bulk orders at the lower bulk critical temperature in the presence of an already ordered
surface. This latter transition is coined extraordinary transition. The possible existence
of a magnetic surface transition at temperatures higher than the bulk critical temperature
has been suggested to exist for various compounds as for example
Gd \cite{Rau87,Tan93,Tob98,Shi00}, Tb \cite{Rau83,Rau88}, FeNi$_3$ \cite{Mam87}, 
NiO \cite{Mar99,El02}, NbSe$_2$ \cite{Mur03} or Ni-Al alloys \cite{Pol95}.
However, the experimental situation is usually not very clear. A good example for the
encountered experimental difficulties is Gd. Whereas it
was believed during many years that Gd 
undergoes a transition to a surface-ordered, bulk-disordered phase
some 80 degrees above
the bulk critical temperature, a recent study \cite{Arn00} claims that for pure Gd surface
and bulk order at the {\it same} temperature, thus giving rise to the ordinary transition
behaviour.

Usually, the surface transition is considered to be of minor theoretical interest.
This is due to the fact that at the
surface transition the surface
critical behaviour of a $d$-dimensional semi-infinite system
is expected to be identical to that of the corresponding ($d-1$)-dimensional bulk system.
Indeed, for perfect surfaces, the critical
exponents of the two-dimensional bulk Ising model are encountered at the
surface transition of the three-dimensional semi-infinite Ising model.
However, the situation is not so simple
for non-perfect surfaces presenting edges or extended surface defects. As will be discussed
in detail in Sections 3 and 4, the local critical behaviour at non-perfect surfaces may
be non-universal at the surface transition. In that case, local critical exponents
which vary continuously
as a function of the local couplings are encountered. Furthermore, the presence of the
disordered bulk is reflected by the values of the local critical exponents.
It is worth noting that these effects are
not restricted to the Ising model but are also encountered in the three-dimensional
easy-axis anisotropic
Heisenberg model whose surface transition belongs to the universality class of
the two-dimensional Ising model.

At the extraordinary transition, the bulk orders in the presence of an ordered surface.
The extraordinary transition with enhanced surface couplings
and vanishing surface field is equivalent to the normal transition where the
bulk orders in the presence of a field acting on the surface.
This has been conjectured by Bray and Moore \cite{Bra77} and later proven by
Burkhardt and Diehl for the Ising model \cite{Bur94}.
The normal transition
occurs in confined binary liquid mixtures at their bulk
critical point. The equivalence of these two transitions is very useful for
experimental studies as surface ordering fields are commonly encountered whereas physical
systems with a genuine extraordinary transition are very scarce. To my knowledge,
the extraordinary transition has only been investigated experimentally for NiO \cite{Mar99,El02}.
It should also be noted that some model systems possess a normal transition without
exhibiting an extraordinary transition. A good example is the two-dimensional
semi-infinite Ising model with short-range interactions.

Surface critical exponents at the extraordinary transition can be completely expressed
by bulk exponents. The singular part of the surface magnetisation, for example, varies
close to the bulk critical temperature as $m_1^{(sing)} \sim \left| T - T_{c}
\right|^{2 - \alpha_b}$.

To conclude this Section, let us finally mention the interesting possibility that
the surface critical behaviour depends on the surface orientation. Examples where this
dependence has been proven are binary alloys with a continuous phase transition \cite{Dre97}
and Ising antiferromagnets in the presence of a magnetic field \cite{Dre97,Abr00}.
Ordinary transition behaviour is encountered in these systems for symmetry-preserving
surface orientations, whereas symmetry-breaking orientations lead to normal critical
behaviour.

\subsection{Thin films}
Thin film magnetism is the subject of intensive current research activities, see 
\cite{Pou99} for a recent review.
For the investigation of magnetic properties of thin films
experimentalists have a large variety of experimental
techniques at their disposal,
ranging from ferromagnetic resonance to magneto-optic Kerr effect measurements.
This has led to a large amount of interesting results concerning the magnetism
of thin films.

Critical phenomena in thin films have also been studied in recent years. In the
following, I discuss some of the more general aspects of thin film critical behaviour,
focusing thereby on thin films with perfect surfaces. The critical behaviour of more
realistic films will be discussed in Section 4.2.

First one has to remark that the critical temperature in thin films 
is a function of the number of layers
forming the film. The temperature shift has been observed in numerous
experimental studies and it has also been investigated extensively in theory
\cite{Ou97,Hen98,Sab00,Zha01,Cab02}.
For thick films with $L$ layers,
the deviation from the bulk critical temperature $T_c(\infty)$
is described by the well-known finite-size scaling relation \cite{Bar83}
\begin{equation} \label{gl:5_1}
1-T_c(L)/T_c(\infty) \propto L^{-\lambda}.
\end{equation}
The shift exponent $\lambda$ is given by $\lambda = 1/\nu_b$ where $\nu_b$ is the 
correlation length
critical exponent of the infinite system.
In the ultrathin film limit, a linear dependence of $T_c(L)/T_c(\infty)$ on the film thickness
is observed, see \cite{Zha01} for a recent discussion.

When analysing critical quantities in films with varying thicknesses, a dimensional
crossover from three-dimensional criticality for thicker films
to two-dimensional critical behaviour for ultrathin films
is observed \cite{Kre91,Kre92a,Kre92b,Li92}, due to the truncation
of the correlation length normal to the film \cite{Nic00}. This crossover has been
studied numerically in Ising films by analysing the critical exponent of the total
film magnetisation \cite{Sch96}. Note that from a puristic point of view, the two-dimensional
critical exponents should be observed for every finite film in a small temperature range
near the critical point. However, as the width of this temperature window decreases
rapidly for increasing film thickness, this temperature range may not be easily accessible
in experiments or in computer simulations. Recent studies investigated this
crossover in various systems using local effective critical exponents \cite{Chu00,Mar00,Mou03}.

In thicker films of some systems,
a crossover from a three-dimensional Heisenberg to a three-dimensional
Ising behaviour may be observed \cite{Li92}. The formation of an easy-magnetisation axis
is due to an increase of the magnetic anisotropy energy.
In fact, the reduced symmetry at surfaces increases the anisotropy energy
as compared to bulk systems where it usually is small. A change of the direction
of the easy axis is often observed when changing the thickness of the film. For example,
for Ni/Cu(001) the easy axis is in-plane for ultrathin films, but an out-of-plane
easy axis is observed for films with more than 7 monolayers \cite{Sch94a}. Long-range
dipolar interactions contributing to the anisotropy are responsible for this behaviour.
A further contribution to the magnetic anisotropy energy has its origin in the fact
that (ultra)thin magnetic films are grown on a substrate. Indeed, a distortion of the lattice
due to strain between the magnetic layers and the substrate may
change the magnetic anisotropy energy as compared to the bulk system. The magnetic anisotropy
energy is the subject of numerous theoretical studies, using different analytic
methods \cite{Guo00} or {\it ab initio} techniques \cite{Hal98}.
In a statistical treatment of surface phenomena,
the change in the magnetic anisotropic energy due to the presence of surfaces is usually
modeled by effective short-range couplings with varying coupling constants, see \cite{Ou97}
for an example. Furthermore, when studying thin film critical behaviour theoretically, 
films are usually
supposed to be free standing.

Most thin films may be grouped into two different universality classes with respect
to their critical behaviour.
Films with an out-of-plane easy axis are theoretically described by Ising
models (e.g.\ the Fe/Ag(100) system \cite{Qiu94}),
whereas films with easy-plane magnetisation are modeled by $XY$-models
(e.g.\ the Ni/Cu(100)
system \cite{Hua94}). Interestingly,
a given material may exhibit both types of magnetisation, depending on the orientation of the
film and the nature of the substrate. For example, Ni(001) grown on Cu(001) presents
an out-of-plane easy axis, whereas for Ni(111) on Re(0001) the magnetisation is in-plane.
Spin reorientation transitions as a function of surface and bulk anisotropies have been
studied theoretically for thin ferromagnetic films as well as for semi-infinite
ferromagnetic systems in \cite{Pop01}.

\subsection{Surface critical behaviour near a Lifshitz point}
Competing interactions are encountered in a large variety of physical systems as,
among others, magnetic systems, alloys or ferroelectrics \cite{Bak82,Sel88,Yeo88,Sel92,Neu98}.
These interactions may lead to
rich phase diagrams with a multitude of commensurately and incommensurately modulated phases as well
as to special multicritical
points called Lifshitz points. At a Lifshitz point, a disordered, a uniformly
ordered and a periodically ordered phase become indistinguishable \cite{Hor75a}.
A large number of systems (as, e.g., magnets, ferroelectric liquid crystals, 
uniaxial
ferroelectrics or block copolymers)
have been shown to possess a Lifshitz point.

From a theoretical point of view the best studied Lifshitz point is that
encountered in the three-dimensional ANNNI model \cite{Ell61,Sel88,Yeo88}.
The Hamiltonian of this model, defined on a simple cubic lattice, reads
\begin{eqnarray}
H = & -&  J \sum\limits_{xyz} S_{xyz} \left( S_{(x+1)yz} + S_{x(y+1)z} \right)
- J  \sum\limits_{xyz} S_{xyz} \, S_{xy(z+1)} \nonumber \\
& +&  \kappa \, J \sum\limits_{xyz} S_{xyz} \, S_{xy(z+2)} \label{gl:6.1}
\end{eqnarray}
with $S_{xyz}= \pm 1$. Here $J > 0$ and $\kappa > 0$ are coupling constants. In the planes nearest
neighbour spins are coupled ferromagnetically with the coupling constant $J$,
whereas in $z$- or axial-direction competition between ferromagnetic
nearest neighbour and antiferromagnetic next-nearest neighbour couplings
takes place, leading to the appearance of spatially modulated phases.
In the infinite system infinitely many commensurately and incommensurately modulated phases
appear in the ($T$,$\kappa$) phase diagram 
\cite{Fis80,Sel84,Sel88}. In thin ANNNI films, however, only modulated phases
compatible with the thickness of the film may be stabilized, leading to a different 
phase diagram for every film thickness \cite{Sel00,Sel02a,Sel02b}.
Note that the related phase transitions in thin helimagnetic and incommensurately
modulated films have also been the subject of recent studies \cite{Mel03,Cha03}

The Lifshitz point encountered in the three-dimensional ANNNI model
is of the uniaxial Ising type, the order parameter at this multicritical
point having only one component. The term uniaxial denotes the fact that wave vector instabilities only show
up in a single direction. In general, the Ginzburg-Landau free energy density in $d$ dimensions may be written
in the presence of a uniaxial Lifshitz point in the following form:
\begin{equation} \label{gl:6.3}
f = a_2 \, \phi^2 + a_4 \, \phi^4 + b_1 \, \left| \nabla_1 \, \phi \right|^2 + b_2 \left| \nabla_{(d-1)}
\, \phi \right|^2 + c_1 \left| \nabla_1^2 \, \phi \right|^2
\end{equation}
where $\phi$ is the order parameter, $\nabla_1$ the space derivative in the axial direction, whereas
$\nabla_{(d-1)}$ is the gradient operator in the directions perpendicular to that direction.
The coefficient $b_1$ 
in equation (\ref{gl:6.3}) may change sign because of the
competition between ferro- and antiferromagnetic interactions 
along the special direction. At the Lifshitz point,
$b_1$ vanishes and the last term in equation (\ref{gl:6.3}) becomes relevant.

A uniaxial Lifshitz point is only a special case of more general Lifshitz points \cite{Hor75a}
(see \cite{Die02a} for a recent brief review) characterized by the
number of space dimensions, $d$, the number of order parameter components, $n$, 
and the dimensionality $m$ of the
subspace where the wave vector instability occurs. The Lifshitz point encountered in
the three-dimensional ANNNI model is then
given by the set $(d,n,m)=(3,1,1)$. 
Uniaxial Lifshitz points are strong anisotropic equilibrium critical points
where the correlation lengths
parallel and perpendicular to the axial axis diverge with different critical
exponents $\nu_\parallel^L$ and $\nu_\perp^L$.
The discovery of the existence of this
kind of multicritical points in 1975 led to numerous theoretical studies of their bulk critical
properties. 
With the exception of an early attempt by Gumbs \cite{Gum86}, surface critical phenomena
at a bulk Lifshitz point
have only been studied very recently \cite{Bin99,Fri00,Jac01,Ple02b,Die03a,Die03b}.
Most of the results have been obtained for the semi-infinite ANNNI model.

\begin{figure}
\centerline{
\psfig{figure=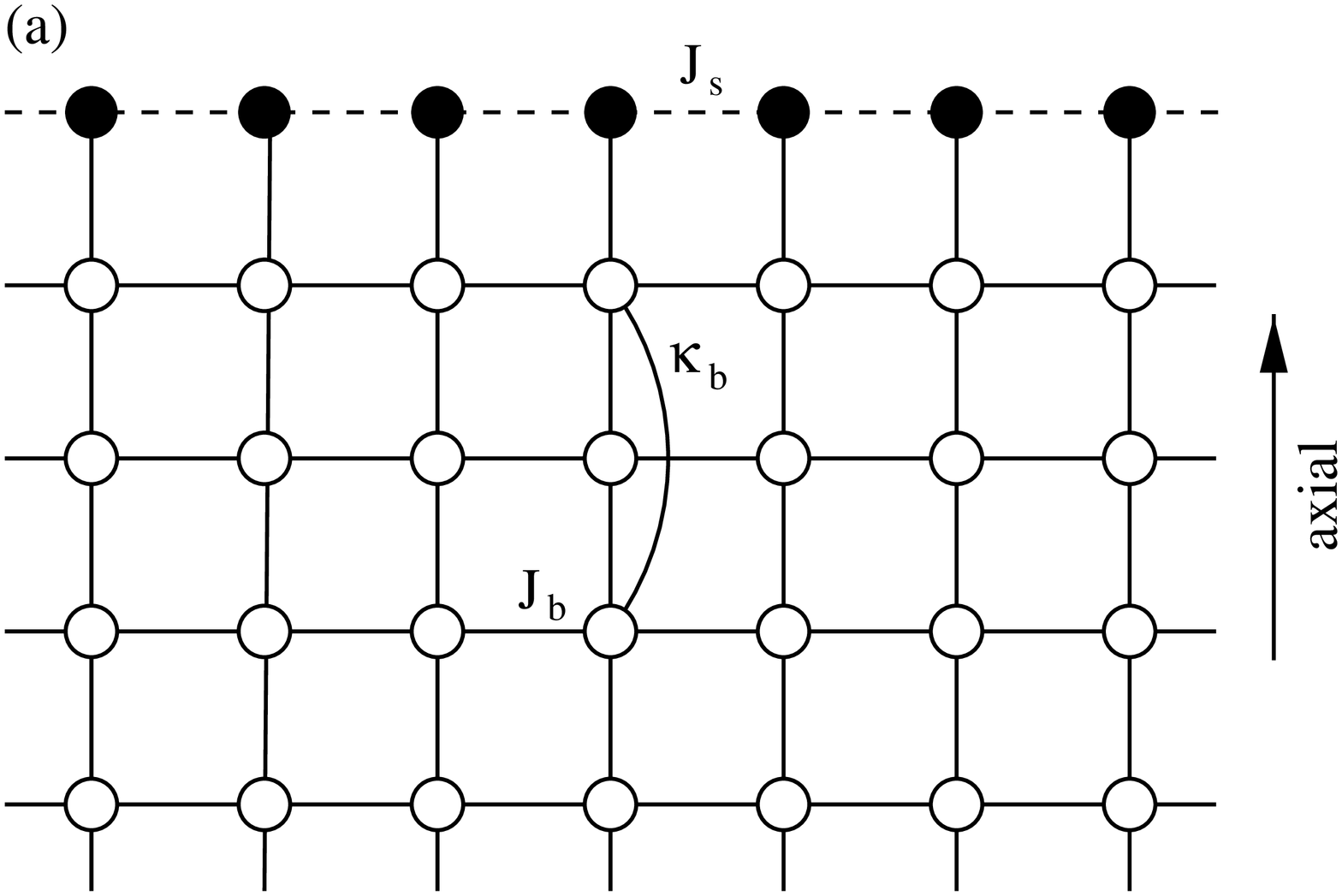,width=4.6cm,angle=270}
\hspace*{0.5cm}
\psfig{figure=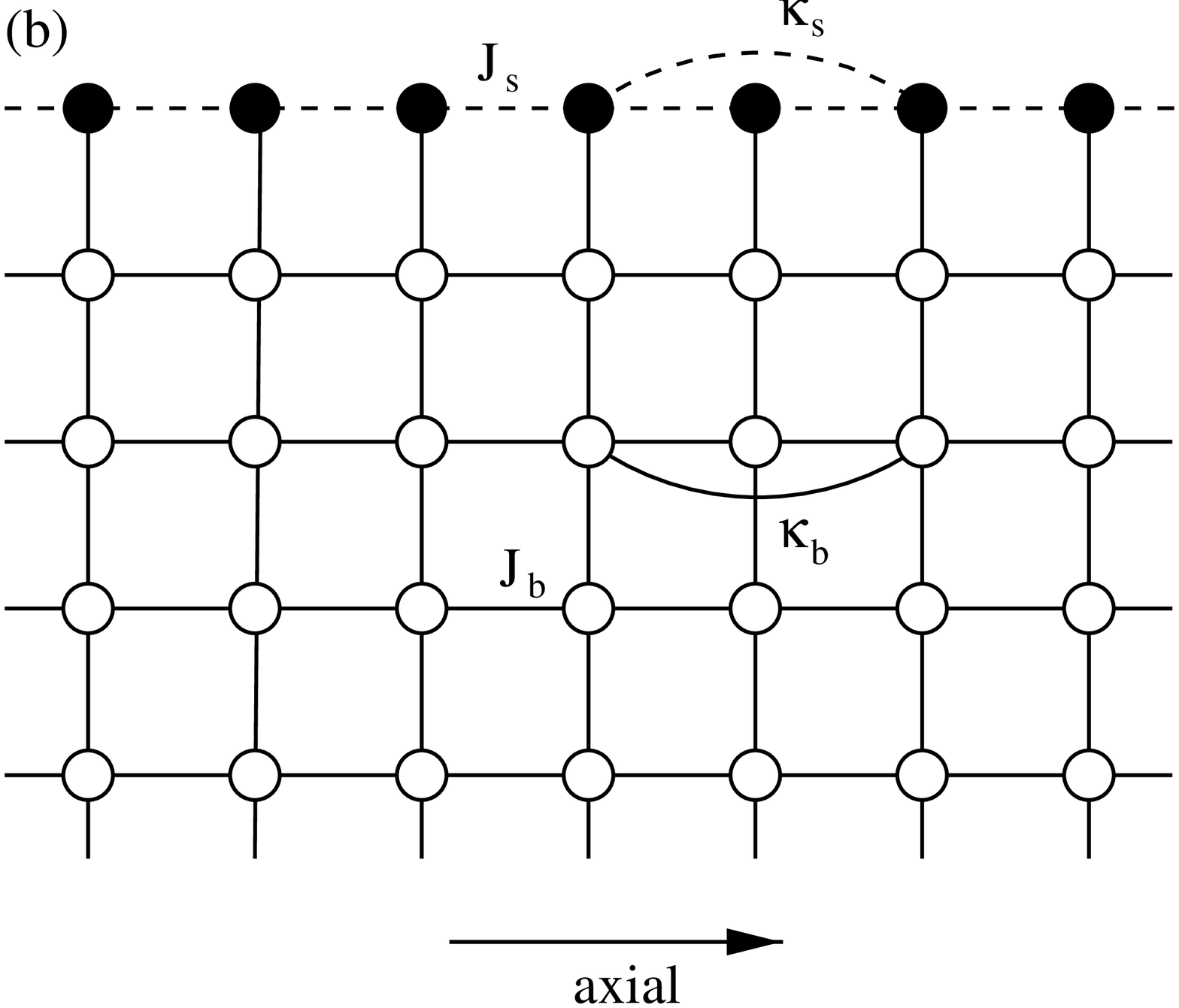,width=5.4cm,angle=270}}
\caption
{\label{fig_annni_1} Cross sections of semi-infinite three-dimensional ANNNI models
showing
two different types of surface orientations:
(a) surfaces perpendicular to the axis of competing interactions, (b)
surfaces parallel to this axis. $J_b$ and $J_s$ denote the nearest neighbour
bulk and surface couplings, respectively, whereas the axial next-nearest
neighbour interactions are labeled by the bulk, $\kappa_b$, and surface, $\kappa_s$,
competing parameters. Surface lattice sites are represented by filled points.
}
\end{figure}

Due to the anisotropy of the ANNNI model, surfaces with different orientations
are not equivalent. The following two surface orientations have been
considered (see Figure \ref{fig_annni_1}): surfaces perpendicular
to the axis of competing interactions (case A) and surfaces parallel to this
axis (case B). 
As usual, the index $b$ ($s$) indicates bulk (surface)
couplings in the following.

For case A modified surface couplings connecting neighbouring surface spins
are introduced in addition to the usual ANNNI interactions, see
Figure \ref{fig_annni_1}a. Three different
scenarios have to be distinguished, depending on the value of the bulk competing
parameter $\kappa_b$. When $\kappa_b$ is smaller than the Lifshitz point
value $\kappa_b^L=0.27$, the bulk undergoes
a second order phase
transition between the disordered high temperature phase and the ordered,
ferromagnetic, low temperature phase at the critical temperature $T_c(\kappa_b)$.
This phase transition belongs to the
universality class of the 3d Ising model. Consequently, the surface phase diagram will
resemble that of the 3d semi-infinite Ising model. At the bulk Lifshitz point, 
$\kappa_b=\kappa_b^L$,
the recent mean-field treatment \cite{Bin99} yields for the semi-infinite ANNNI model
a surface phase diagram similar to the Ising model, but with a set of critical
exponents different from those of the Ising model. The predicted
existence at the bulk Lifshitz point \cite{Bin99}
of a special transition point is in variance
with the earlier treatment of \cite{Gum86} but agrees with
the Monte Carlo results of \cite{Ple02b}.
Finally, for
axial next-nearest neighbour bulk couplings $\kappa_b > \kappa_b^L$ the bulk
phase transition from the disordered to the modulated phase
belongs to the universality class of the 3D $XY$ model \cite{Gar76}.
It may then be argued that at the ordinary transition one is dealing with a critical
semi-infinite 3D $XY$ model with Ising-like surface exchange anisotropies. Surface exchange
anisotropies being irrelevant near the ordinary transition \cite{Die84}, one therefore
expects the ordinary transition critical behaviour of the semi-infinite 3D ANNNI
model with $\kappa_b > \kappa_b^L$ to be identical to that of the three-dimensional semi-infinite
$XY$ model \cite{Lan89}. A surface transition belonging to the 3D Ising universality class
will again be encountered for strong surface enhancements. The ordinary transition and
the surface transition lines are then expected to meet at an anisotropic special
transition point \cite{Die84}.

Case B consists of surfaces oriented parallel to the axial direction (see Figure 
\ref{fig_annni_1}b).
The introduction of modified
nearest neighbour, $J_s$, {\it and} axial next-nearest neighbour couplings,
$\kappa_s > 0$, in the surface layer leads to intriguing and very
complex situations. First, one has to note that the critical value of the
coupling ratio $r_{sp}$, needed for the surface to get critical by itself,
depends both on the bulk, $\kappa_b$, and on the surface, $\kappa_s$, competing parameters.
Three different cases may be distinguished,
depending on the value of $r=J_s/J_b$.

\begin{itemize}
\item[1)] For $r < r_{sp}(\kappa_b,\kappa_s)$ the ordinary surface transition is encountered,
with an Ising bulk ordering ($\kappa_b < \kappa_b^L$), a Lifshitz point bulk ordering
($\kappa_b = \kappa_b^L$), or a modulated bulk ordering ($\kappa_b > \kappa_b^L$).
\item[2)]  For $r > r_{sp}(\kappa_b,\kappa_s)$ the surface orders at a higher temperature
than the bulk. For $\kappa_s < 1/2$ the surface transition belongs to the universality
class of the two-dimensional Ising model, whereas for $\kappa_s > 1/2$ a floating incommensurate
phase appears in the surface layer. No surface transition, where the surface alone orders, will be
encountered for $\kappa_s = 1/2$, as in the 2d ANNNI model the paramagnetic phase extends down
to $T=0$ for this value of $\kappa_s$ \cite{Sel88}.
\item[3)] For $r = r_{sp}(\kappa_b,\kappa_s)$ the special surface transition point is encountered
where the ordinary transition line and the surface transition line merge. Depending on the
values of $\kappa_b$ and $\kappa_s$, very interesting possibilities arise. Whereas for
$\kappa_b < \kappa_b^L$ and $\kappa_s < 1/2$ the usual scenario of an Ising ordinary transition
meeting an Ising surface transition is encountered, for $\kappa_b = \kappa_b^L$ and $\kappa_s < 1/2$
a Lifshitz point ordinary transition merges with an Ising surface transition.
These are the only cases studied so far in some detail \cite{Fri00,Ple02b}.
However, more exotic multicritical points may also be encountered. For example, one may
have the case that at the special transition point an ordinary transition line with a modulated
bulk ($\kappa_b > \kappa_b^L$) meets a surface transition line to a floating incommensurate
phase in the surface layer ($\kappa_s > 1/2$). Finally, I should also mention the rather academic
possibility that in four dimensions a Lifshitz point ordinary transition ($\kappa_b = \kappa_b^L$)
merges with a Lifshitz point surface transition ($\kappa_s = \kappa_s^L$).
\end{itemize}

In their mean-field treatment Binder, Frisch, and Kimball \cite{Bin99,Fri00} 
considered surfaces oriented either perpendicular \cite{Bin99}
or parallel \cite{Fri00} to the axis of competing interactions. For both cases
the mean-field surface critical exponents at the Lifshitz point
were determined at the ordinary transition and two different
sets of critical exponents were obtained, see Table \ref{table_annni_1}.
This dependence of the ordinary transition critical exponents on the surface
orientation is explained by the fact that the bulk Lifshitz point is a
strongly anisotropic equilibrium critical point. 

\begin{table}
\begin{tabular}{|l||l|l|l|l|l|l|}
\hline
& & $\beta_1^L$ & $\gamma_1^L$ & $\gamma_{11}^L$ & $\beta_s^L$ & $\gamma_s^L$  
\\ \hline\hline
case A & OT/MF \cite{Bin99} & 1 & $ 1/2 $ & $-1/4$& 1/4 & 5/4 \\
\hline
& OT/MC \cite{Ple02b} & $0.62(1)$ & 0.84(5) & $-0.06(2)$ & $-0.14(4)$ & 1.69(7) \\
\hline
& SP/MC \cite{Ple02b} & 0.22(2) & 1.28(8) & 0.76(5) & &  \\
\hline\hline
case B & OT/MF \cite{Fri00} & 1  & 1/2 & $- 1/2$ & 0  & 3/2 \\
\hline
& OT/MC \cite{Ple02b} & $0.687(5)$ & 0.82(4) & $-0.29(6)$ & $-0.46(3)$ & 1.98(8)  \\
\hline
& OT/RNG \cite{Die03b} & 0.697 & 0.947 & $-0.212$ & $-0.462$ & 2.106 \\
\hline
& SP/MC \cite{Ple02b} & 0.23(1) & 1.30(6) & 0.72(4) & &  \\
\hline
\end{tabular}
\caption
{\label{table_annni_1} Surface critical exponents at a uniaxial bulk Lifshitz point
with the surface layer oriented
perpendicular (case A) and parallel (case B) to the axial direction.
OT: ordinary transition, SP: special transition point, MF: mean-field theory, MC: Monte Carlo
simulations, RNG: renormalization group methods.
}
\end{table}

One should note that various scaling relations are fulfilled for case A \cite{Bin99}, as for example
\begin{eqnarray}
\gamma_s^L & = & \gamma_b^L + \nu_\|^L \label{gl:6.17} \\
\beta_s^L & = & \beta_b^L - \nu_\|^L \label{gl:6.18} \\
\gamma_s^L & = & 2 \gamma_1^L - \gamma_{11}^L \label{gl:6.19}
\end{eqnarray}
Here $\beta_b^L$, $\gamma_b^L$, and $\nu_\|^L$ are Lifshitz point bulk
critical exponents which take in mean-field approximation the values
$\beta_b^L=1/2$, $\gamma_b^L=1$, and $\nu_\|^L=1/4$.
Whereas (\ref{gl:6.19}) also holds for case B,
the scaling relations (\ref{gl:6.17}) and (\ref{gl:6.18}) have to be modified in
this case. Indeed, the behaviour of excess quantities is governed
close to a bulk critical point by the bulk correlation length along
the direction perpendicular to the surface. The Lifshitz point being
an anisotropic critical point characterized by two
correlation lengths diverging with different critical exponents,
$\nu_\perp^L$ should be used in case B instead of $\nu_\|^L$. This then leads
to the scaling relations
\begin{equation} \label{gl:6.20}
\beta_s^L=\beta_b^L - \nu_\perp^L
\end{equation}
and
\begin{equation}\label{gl:6.21}
\gamma_s^L=\gamma_b^L+\nu_\perp^L.
\end{equation}

Recent Monte Carlo simulations \cite{Ple02b} revealed that the predictions of 
mean-field theory \cite{Bin99,Fri00} are qualitatively correct. 
As shown in Table \ref{table_annni_1} the values of the surface critical exponents
at the ordinary transition in vicinity of the Lifshitz point indeed depend
on the surface orientation. It is worth noting that for both surface orientations
the value $\beta_1^L$ for the surface order parameter is clearly smaller
than the corresponding value obtained in the semi-infinite Ising model. Mean-field
theory yields for all these cases the same value $\beta_1^{MF}=1$.

Estimates for the Lifshitz point surface critical behaviour at the special transition
point have also been included in
Table \ref{table_annni_1}. Interestingly, the values of the critical exponents are
very similar for the two different surface orientations. 
This is a strong indication that at the bulk Lifshitz point
the surface critical behaviour at the special transition point may not depend
on the orientation of the surface with respect to the axial axis.
One may relate this observation to the fact that
both the bulk (diverging bulk correlation length) and the surface (diverging correlation length
for correlations along the surface layer) are critical at the special transition point. It may then
be argued that the surface critical behaviour is governed to a large extend by the critical
fluctuations along the surface, so that the surface orientation
with respect to the direction of competing interactions
is only of minor importance. At the ordinary transition, however, the surface is not
critical and the surface critical behaviour is governed exclusively by the critical bulk fluctuations,
leading to orientation dependent
critical exponents because of the anisotropic scaling at the bulk Lifshitz point.

Very recently Diehl and coworkers \cite{Die03a,Die03b} analysed the surface
critical behaviour at bulk Lifshitz points using renormalization group methods. 
They thereby considered general $m$-axial Lifshitz points where the wave vector
instability takes place in an $m$-dimensional subspace of the $d$-dimensional space.
Thus for $m=1$ one recovers the situation encountered in the ANNNI model.
Restricting themselves to surfaces parallel to the modulation axes (i.e. to case B),
they constructed the appropriate continuum $\left| \phi \right|^2$ models and computed
the critical exponents at the ordinary transition to order $\varepsilon^2$. Their results
for $m=1$ are included in Table \ref{table_annni_1}. One notes a very good agreement
with the Monte Carlo results obtained in \cite{Ple02b}.

\subsection{Critical dynamics at surfaces}
Besides changing the local static critical behaviour surfaces have also an effect on
the dynamic critical behaviour \cite{Kum76,Die83,Kik85,Rie85,Xio89a,Xio89b,Fra89,
Bin91,Die92,Die94a,Wic95,Rit95,Maj96,Muk99,Kre01,Die02,Ple03}. 
A central aspect of the works on dynamic surface critical behaviour 
concerns the possible classification of the distinct surface dynamic universality
classes, similar to what has been done in the past for the dynamical bulk
critical behaviour \cite{Hoh77}. In that context semi-infinite extensions
of the well-known bulk stochastic models are considered. Interestingly, different
surface dynamic universality classes may be encountered for a given bulk
model. This has especially been studied in the semi-infinite extension of the relaxation 
model $B$ where the bulk order parameter is conserved \cite{Die94a,Wic95}.
It has been shown that the presence of nonconservative surface terms, leading
to a nonconserved local order parameter in the vicinity of the surface, yields
a different dynamic critical universality class as compared to the case
where these terms are absent. These two universality classes share the same
critical exponents but are characterised by different scaling functions
of dynamic surface susceptibilities.

It is important to note that no genuine dynamic surface exponent exists \cite{Die83}.
Indeed all exponents describing the equilibrium critical behaviour of dynamic quantities can be
expressed entirely in terms of static bulk and surface exponents and the dynamic
bulk exponent $z$. For instance the dynamic spin-spin autocorrelation function decays
for long times as $t^{-2 x_1/ z}$ where $x_1$ is the surface scaling dimension. The
value of the dynamic
exponent $z$ is approximately 2.17 (2.04) in the two-dimensional (three-dimensional)
Ising model. 

The effect of surfaces on {\it nonequilibrium} dynamics after a quench from high
temperatures, $T \gg T_c$, to the critical temperature has been investigated in
\cite{Rit95,Maj96,Ple03}. These studies are to some extent complementary to the
investigations of universal short-time behaviour in the bulk \cite{Jan89}.
Indeed, similar to the bulk magnetisation, the surface magnetisation 
displays at early times a power-law behaviour
\begin{equation} \label{eq_dyn_1}
m_1(t) \sim m_{1,0} \, t^\theta
\end{equation}
with $\theta=( x_i - x_1)/z$. Here $m_{1,0}$ is the small surface magnetisation of the 
initial state, 
$x_1$ is the scaling
dimension of the surface magnetisation, whereas
the nonequilibrium exponent $x_i$ is the scaling dimension
of the initial magnetisation \cite{Jan89}.
A corresponding power-law behaviour is also obtained for the local nonequilibrium
autocorrelation in the long time limit $t \gg 1$ \cite{Rit95,Maj96}. 
It has been revealed recently \cite{Ple03} that in the case $x_i < x_1$ a new effect, called
cluster dissolution, takes place, which leads to an unconventional, stretched exponential
dependence of the short-time autocorrelation. 
A crossover to a power-law behaviour is then observed at later
times. Interestingly, this stretched exponential behaviour is observed in the important
case of the three-dimensional semi-infinite
Ising model where $x_i = 0.53$ and $x_1 = 1.26$ \cite{Ple03}.

There has been some progress in the analysis of dynamics in semi-infinite systems,
but the situation is still very unsatisfactory as the effects of surfaces on dynamics
have up to now only been studied in a very unsystematic way. From the experimental point
of view the situation is even worse as no experimental studies of dynamic surface critical behaviour
have yet been published. It has to be noted that some possible experiments,
involving inelastic magnetic scattering of neutrons \cite{Die85} or M\"{o}ssbauer and 
NMR spectra \cite{Die90},
have been discussed in the literature, but they have not yet been realised.

\subsection{Surfaces and fields}
Up to this point we have only discussed surface phase diagrams in absence of
external fields. In fact, the global phase diagram 
not only depends on temperature and on surface enhancement, but also on 
symmetry-breaking bulk and surface fields \cite{Nak82}. A discussion of the rich field
of wetting phenomena, including for example critical wetting or prewetting,
is beyond the scope of this article, even so some recent investigations
of wetting criticality in wedge-shaped geometries will briefly be mentioned
in Section 3.3. The reader interested in this field is referred to the review
by Dietrich \cite{Die88} and to the overview of experiments by Bonn and
Ross \cite{Bon01}. I also refrain from dealing with the related topic
of localisation-delocalisation transitions observed for example in Ising films
with competing surface fields. This topic was the subject of a recent review
article by Binder {\it et al.} \cite{Bin03}.

Instead, I focus here on the normal transition and, especially, on the
crossover between ordinary and normal transitions in presence of weak surface
fields. This crossover has attracted much interest in the past and is also
of relevance when discussing recent experimental studies of surface critical
behaviour.

At the normal transition the bulk orders at the bulk critical temperature in
presence of an ordered surface. The non-zero surface magnetisation is thereby generated
by a symmetry-breaking surface field. A similar situation is encountered
at the extraordinary transition where the finite surface magnetisation at the
bulk critical temperature is due to strong enhancement of the surface couplings.
This two different transitions have been shown \cite{Bra77,Bur94} to belong to
the same surface universality class. 

It seems natural to expect in the vicinity of the normal transition a monotonic 
decaying magnetisation profile for $T \geq T_c$. This is indeed observed for strong
surface fields, but for weak fields a more complex behaviour with a non-monotonic
profile may be observed. Assuming that the surface enhancement $c$ is subcritical, i.e.\
that without external fields one would be at the ordinary transition, 
it has been shown \cite{Sym81,McA93,Rit96} that a small surface field leads to a
short-distance increase of $m(z)$. This critical adsorption in systems
with weak surface fields has been studied subsequently
in a series of papers \cite{Cze97a,Cze97b,Cia97,Mac99,Mac03}.
The increase of the layer magnetisation is due to the fact that a weak surface field
gives rise to a macroscopic length scale, yielding a power law behaviour
\begin{equation} \label{eq:weak}
m(z) \sim h_1 \, z^\kappa ~~~ \mbox{with} ~~~ \kappa= d-1 -x_b -x_1
\end{equation}
where $x_b$ and $x_1$ are the bulk and surface scaling dimensions, respectively.
This increase continues up to a distance $l^{ord} \sim h^{1/(d-1-x_1)}$, where a
crossover to the normal monotonic decrease $m(z) \sim z^{-x_b}$ sets in. Recent studies
of critical adsorption in a weak surface field for a homologous series of critical
liquid mixtures \cite{Cho01} have permitted to observe the predicted increase experimentally.
It is worth noting that for surface enhancement corresponding to the special transition
point $m(z)$ displays in the weak surface field limit a monotonic decay, but with two
different power laws in the limits $z \longrightarrow 0$ and $z \longrightarrow \infty$ \cite{Bre83}.

It has been suggested \cite{Rit96,Rit98} that 
this crossover between the ordinary and the normal transitions may explain some
puzzling experiments on surface critical behaviour. In these experiments \cite{Mai90,Kri97}
critical exponents compatible with the ordinary transition critical behaviour were
measured, but at the same time the existence of residual long-range surface order was
revealed at temperatures above the bulk critical temperature. This behaviour is readily
explained by assuming that a weak surface field exists, yielding a surface structure factor
governed by the ordinary behaviour in the case that $l^{ord}$ exceeds the bulk correlation
length. This interpretation assumes the existence of a surface field. It is therefore
important to note 
that this kind of weak surface field can indeed arise in the studied compounds
due to non-ideal stoichiometry effects \cite{Sch93,Lei98,Dre97}.

\section{Surfaces with edges and corners}
The semi-infinite model with a flat surface may be considered to be a special
case of a more complex wedge geometry where two planes meeting at an angle $\theta$
form an infinite edge. For $\theta= \pi$ the flat surface is recovered. Cardy 
showed that at the ordinary transition edge critical exponents depending
continuously on $\theta$ arise on purely geometrical grounds \cite{Car83}. For a given opening angle
$\theta$, however, the values of the critical exponents are expected to be universal and independent
of microscopic details such as the strengths of the coupling constants or the lattice type.
Whereas edge singularities at the ordinary transition have been studied intensively,
especially in two dimensions \cite{Igl93}, edge critical
behaviour at the surface transition and at the normal transition have in general been overlooked.
The recent investigations
of edges and corners in systems with enhanced surface couplings and/or
surface fields revealed some unexpected phenomena which will be reviewed in Sections 3.2 and 3.3.

\subsection{Ordinary transition}
Cardy considered in his work $d$-dimensional $O(n)$ models containing edges formed by
$(d-1)$-dimensional hyperplanes
meeting at an angle $\theta$ \cite{Car83}. For this geometry local critical exponents 
changing continuously with
the angle
$\theta$ are already obtained in mean-field approximation. The edge magnetisation critical   
exponent for example is given by
\begin{equation} \label{gl:4_1}
\beta_2^{MF}= \frac{1}{2} + \frac{\pi}{2 \theta}.
\end{equation}
For the opening angle $\theta=\pi$, the surface critical exponents are recovered. 
In the following, we use the
same notation as Cardy and refer to edge quantities by the subscript 2. Corner quantities resulting, in
dimensions $d \geq 3$, from the meeting of three hyperplanes will carry the subscript 3.

Near the bulk critical point $T_c$, the edge energy density has the scaling form \cite{Car83}
\begin{equation} \label{gl:4_2}
f_e = |t|^{(d-2)\nu_b} \, \psi \left( h |t|^{- \Delta_b}, h_1 |t|^{- \Delta_1}, 
h_2 |t|^{-\Delta_2} \right)
\end{equation}
with $t=(T_c-T)/T_c$. $\nu_b$ is the critical exponent of the bulk correlation length,
$h$, $h_1$, and $h_2$ are magnetic fields.
$h_2$ only acts on edge spins, whereas $h$ acts on all spins and $h_1$ on all surface spins.
$\Delta_b$ and $\Delta_1$ are the bulk and surface exponents
appearing in the singular part of the free energy
density of a semi-infinite system.
The new exponent $\Delta_2$, which has been computed in first order
of an $\epsilon=d-4$ expansion in \cite{Car83}, changes continuously with the wedge angle $\theta$.
All edge critical exponents may be expressed by $\Delta_2$ together with bulk and surface critical   
exponents. From the renormalization group point of view, the angle dependence has its origin in
the invariance of the edge under rescaling. This makes the opening angle a marginal variable
and may therefore lead to angle dependent local critical exponents.

Inspired by these results, subsequent work mainly studied the influence of "edges" in two dimensions.
In the following only some selected results obtained in two dimensions are discussed, the
interested reader is referred to the excellent review of Igl\'{o}i, Peschel and Turban \cite{Igl93}
for a more complete account. The "edge" then reduces 
to one point and forms a corner in a two-dimensional system. Based on
numerical and analytical calculations of isotropic two-dimensional Ising models \cite{Bar84},
the critical exponent of the local edge magnetisation was postulated to be
\begin{equation} \label{gl:4_3}
\beta_2= \frac{\pi}{2 \theta}
\end{equation}
for 2d Ising models.
A similar equation was proposed for anisotropic lattices, with the opening angle replaced by an effective
angle depending on the ratio of the different couplings. Using the conformal transformation
$w = z^{\theta/\pi}$, which transforms the semi-infinite system in the $z$-plane to a wedge with
opening angle $\theta$ in the $w$-plane, a simple relation between edge and surface critical
exponents in two dimensions was derived \cite{Car84a,Bar84}. Thus, for the local magnetisation one obtains
\begin{equation} \label{gl:4_4}
\beta_2 = \frac{\pi}{\theta} \, \beta_1
\end{equation}
which is in accordance with equation (\ref{gl:4_3}) for the Ising model. Other work focused on the
temperature behaviour of the local magnetisation in two-dimensional Ising models with various opening
angles and lattice types \cite{Pes85, Kai89, Dav91, Dav97}. Up to now a complete solution has only
been obtained for the square lattice Ising model with $\theta= \pi/2$ \cite{Abr94, Abr95, Abr96}.
Further studies also investigated the influence of edges in other two-dimensional
systems. Examples are polymers \cite{Gut84, Dup86}, regarded as $O(n)$ models
in the limit $n \longrightarrow 0$,
or Potts models \cite{Kar97a}.

A qualitatively different critical behaviour is observed in systems with parabolic
shapes \cite{Pes91, Igl93}. These systems have the remarkable property that they are
asymptotically narrower than wedges.
In this
geometry one does not observe the usual power laws at criticality but stretched
exponentials. Close to the critical point the tip magnetisation
vanishes like $\exp ( - a \, t^{-b} )$ with $a$, $b > 0$ and $t = (T_c-T)/T_c$. 

For small three-dimensional systems, edges and corners may be expected to play a dominant role,
for instance, in nanostructured materials. Studies to reproduce average properties
of such small clusters of atoms have been performed. These include several Monte Carlo
investigations (see, e.g., \cite{Lan76,Mer91,Agu95}). Critical phenomena at edges in
three-dimensional models have, however, been addressed only rarely. Most investigations of
edge criticality
focused on Ising models, but some studies of percolation \cite{Sax87a,Gra92}
and of polymers \cite{Gut84, Gau90} in three-dimensional wedges
have also been published.
Whereas most of these studies of edge critical properties investigated the case that at the bulk critical
point both surfaces forming the edge undergo an ordinary transition, Wang {\it et al.} \cite{Wan90a}
also computed critical exponents for the cases that at least one of
the surfaces undergoes a special transition.
In this renormalization group study, the authors limited themselves to the opening
angle $\theta=\pi/2$. They also studied the corner critical behaviour of cubic systems,
each surface having von Neumann (special transition) or Dirichlet (ordinary transition)
boundary conditions.
Saxena \cite{Sax87b} discussed, based on renormalization group calculations, the possible phase
diagrams of
Ising models with an edge formed by two perpendicular surfaces. Similar methods were used by
Larsson \cite{Lar86}
in his study of Ising edge critical behaviour, limiting himself mainly to the ordinary transition.
The edge exponent $\Delta_2(\theta)$, see equation (\ref{gl:4_2}), was computed for some angles $\theta$ 
by Guttmann and Torrie \cite{Gut84} using high
temperature series expansions. Based on these results, they proposed an expression 
for $\Delta_2(\theta)$ supposed
to be valid for all $\theta$. Early Monte Carlo simulations
of Ising models were done by Mon and co-workers in order to
compute the edge exponent $\Delta_2(\theta=\pi/2)$ \cite{Mon89} as well as
the corresponding corner quantity $\Delta_3$ for a cube \cite{Lai89a}. In these studies, only rather 
small systems were investigated. The critical free energies of Ising models
with edges and corners were also studied \cite{Pri88, Lai89b}, following a similar investigation
in two-dimensional critical systems with corners \cite{Car88}.
$O(n)$ models with edges and corners
were studied in the limit $n \longrightarrow \infty$ in \cite{Kac01}.
The temperature dependence of the edge and corner magnetisations as well as
related quantities were only investigated recently for three-dimensional Ising models in the case
of equal surface and bulk couplings \cite{Ple98b,Ple02a}, using modern
simulation methods.

\begin{figure} 
\centerline{
\psfig{figure=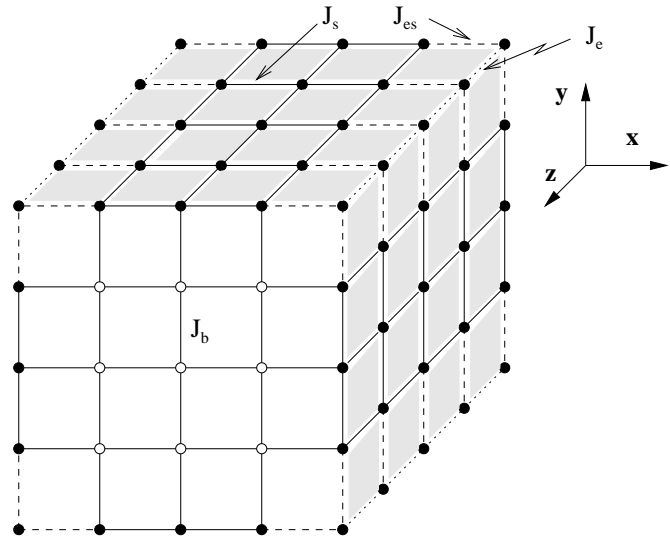,width=7cm,angle=270}
}
\caption
{\label{fig_edge_1} Geometry of a model with (100) and (010) surfaces, i.e.\ edges
with opening angles $\theta = \pi/2$. $J_b$ and $J_s$ are the bulk and surface
couplings, respectively. $J_e$ is the coupling between neighbouring edge spins,
$J_{es}$ the coupling between an edge spin and its neighbouring surface spin.
}
\end{figure}

\begin{figure} 
\centerline{
\psfig{figure=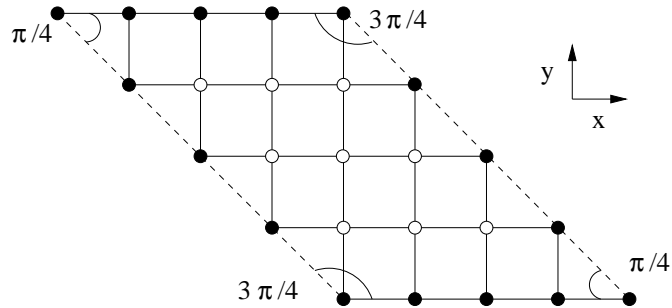,width=4cm,angle=270}
}
\caption
{\label{fig_edge_2} Geometry of a model with edges with opening angles $\theta=\pi/4$ and $\theta=3 \pi/4$.
Shown is a cut through the crystal perpendicular to the edge direction. The full lines
show the couplings between
neighbouring spins, whereas the (110) surfaces are indicated by the dashed lines.
}
\end{figure}

To introduce edges in Ising magnets, periodic boundary conditions along one axis,
the $z$-axis, are applied. The remaining four free surfaces of the crystal may be
oriented in various ways leading to different opening angles $\theta$ at the edges.
As shown in Figure \ref{fig_edge_1}, pairs of (100) and (010) surfaces lead to four equivalent edges
with opening angles $\theta= \pi/2$. The intersections of (100) and (110) surfaces form
two pairs of edges with $\theta=\pi/4$ and $\theta= 3 \pi/4$. This is illustrated in Figure 
\ref{fig_edge_2}.
Besides the bulk coupling $J_b$ and the surface coupling $J_s$, further couplings may
be introduced \cite{Ple98b,Ple99}: two neighbouring edge spins interact with the edge coupling
$J_e$, whereas an edge spin is coupled to its neighbouring surface spin by the edge-surface
interaction $J_{es}$. When studying edge behaviour near the
ordinary transition all couplings can be considered
to have the same strength: $J_e=J_{es}=J_s=J_b$.

\begin{table}
\begin{tabular}{|c||c|c|c|c|}
\hline
 & $\pi/4$ & $\pi/2$ & $3\pi/4$ & $\pi$ \\
\hline\hline
MF \cite{Car83} & 2.50 & 1.50 & 1.17 & 1.00 \\
\hline
RNG \cite{Car83} & 2.48 & 1.39 & 1.02 & 0.84 \\
\hline
HTS \cite{Gut84} & 2.30 & 1.31 & 0.98 & 0.81 \\
\hline
MC \cite{Mon89} & & $1.38(6)$ & & \\
\hline
MC  \cite{Ple98b}  & ~~ $2.3(1)$ & ~~ $1.28(4)$ ~~ & ~~ $0.96(2)$~~  &~~  $0.80(1)$
~~  \\
\hline
\end{tabular}
\caption
{\label{table_edge_1} 
Predictions for the Ising edge magnetisation critical exponents from various methods for four different
opening angles of the three-dimensional wedge. MF: mean-field
approximation, RNG: renormalization group theory, HTS: high temperature series expansions, MC:
Monte Carlo simulations.}
\end{table}

The values of the local critical exponent $\beta_2$ of the edge magnetisation obtained in various 
studies of Ising models are compiled in Table \ref{table_edge_1}. The expected angle dependence of
the local critical exponents is thereby nicely illustrated. 
The values given by mean-field theory are systematically too high,
according to the predictions of the renormalization group \cite{Car83}. These predictions, in turn,
seem to be systematically too large, as suggested both by high temperature series expansions \cite{Gut84}
and by Monte Carlo simulations \cite{Ple98b}. The last two methods yield results which
are in close agreement with each other. The value of $\beta_2$ for $\theta= \pi/2$ derived
from the Monte Carlo study of Mon and Vall\'{e}s \cite{Mon89} differs significantly from
the value obtained from the recent studies using cluster algorithm \cite{Ple98b}, 
presumably due to the small systems investigated  
in that early study.

In order to study the universality of the computed critical exponents, one may investigate,
at a fixed opening angle,
wedges which are formed by different pairs of surfaces.
For example, comparing $\theta=\pi/2$
edges formed by (100) and (010) surfaces or (110) and ($1\overline{1}0$) surfaces show that,
on the one hand, the local magnetisations differ, but that, on the other hand,
the values of the local critical exponents are invariant against rotation of the crystal \cite{Ple98b}.
Note that the invariance of boundary critical exponents against rotation of the crystal
has been discussed and partly even proven for edges in two-dimensional Ising models \cite{Pes85,Igl93}.
Similarly, modifying the strength of the couplings does not alter
the values of the critical exponents at the ordinary transition \cite{Ple98b}. Furthermore,
changing the lattice type is not supposed to change the critical behaviour either.
The independence of the values of the edge critical exponents on the lattice type has been
studied in two dimensions \cite{Igl93}.

Investigations of critical phenomena near corners in three-dimensional systems have up to now
been limited to the special case where the three surfaces forming the corner are mutually
perpendicular. In Monte Carlo simulations, Ising cubes on simple cubic lattices with free surfaces
are considered \cite{Lai89a,Ple98b,Ple00}. The corner magnetisation critical exponent
is thereby determined to be $\beta_3=1.77(5)$ \cite{Ple00}. This value is significantly lower
than the mean-field value $\beta^{MF}_3= 2$ \cite{Igl93}. Analytical results beyond mean-field
are not yet available.

Recent simulations of water in hydrophobic pores also illustrated the effect of curved
surfaces on the values of local critical exponents \cite{Bro04}. In that study Brovchenko {\it et al.}
observed a sharp crossover for cylindrical surfaces, yielding an asymptotic
value for the local order parameter larger than the value $\beta_1 \approx 0.82$ measured
for slitlike pores. 
The critical behaviour characteristic of the ordinary transition 
is expected to be observed in this case not too close to $T_c$ as one is dealing with
the liquid-vapour transition near a weakly attractive surface, see the discussion in Section 2.6.
Of course, the geometry is slightly different to the wedge-shaped geometries discussed
so far in the literature, but nevertheless the kind of systems discussed in \cite{Bro04} 
seems to be a good candidate to study the effect of (generalised) edges on the local
critical behaviour experimentally in the future.

\subsection{Surface transition}
Edge and corner critical properties at the surface transition have been usually overlooked. This problem
was briefly addressed in \cite{Lar86} and \cite{Sax87b}, but no systematic study has been done
until recently \cite{Ple99,Ple00}. As shown in the following, intriguing
phenomena are observed at edges and corners when the surface orders whereas the bulk remains
disordered. Especially, non-universal local critical exponents are encountered at the surface
transition of crystals with edges and corners. As discussed in the Section 3.1,
edge and corner critical exponents at the ordinary transition depend on the opening angle but do not depend
on microscopic details of the model as for example the values of the interactions, the lattice
type or the orientation of the surfaces. The edge and corner behaviour at the surface transition 
is in marked
contrast to this, as, for a fixed opening angle, local critical exponents change continuously
with the strengths of the different couplings \cite{Ple99,Ple00}. As we shall see, the values
of edge and corner critical exponents at the surface transition also reflect the existence of
the disordered bulk.

At the surface transition of the three-dimensional semi-infinite Ising model with
a perfect surface, the critical fluctuations are essentially
two-dimensional. The surface critical exponents reflect this reduced dimensionality, e.g.\
$\beta_1=\beta_{2d}=1/8$. On the other hand, the edge presents a local perturbation, acting
presumably like a line defect in a two-dimensional system. Simple one-dimensional defects
in the two-dimensional Ising model have been studied intensively
\cite{Fis67,Bar79,McC80,Ko85,Tur85,Del94,Osh96,Sim98,Sza99}. It was shown that in
the vicinity of these defects the local magnetic critical exponents are non-universal \cite{Bar79}.
As these exact results
provide a useful framework for discussing the numerical findings of \cite{Ple99} and 
\cite{Ple00}, I briefly
review in the following the main results concerning the local critical behaviour near defect lines
in two-dimensional Ising models.

The plane Ising model with a defect line \cite{Fis67} is an interesting system displaying non-universal
magnetic critical exponents. A simple analysis on the relevance of perturbations shows that in
two-dimensional Ising models an energy-like one-dimensional perturbation is marginal, yielding
continuously varying local critical exponents. This is a consequence of the fact that in the unperturbed
system the correlation length critical exponent $\nu_b$ equals 1 and the scaling dimension of
the surface magnetisation operator $x_1$ is equal to $1/2$ \cite{Igl93}. The non-universal behaviour
was proven for the first time by Bariev \cite{Bar79}. He analysed two types of defect lines:
chain defects, where a column of perturbed couplings with strength $J_{ch}$
is considered, and ladder defects, where modified couplings of strength $J_l$ connect spins 
belonging to two neighbouring columns. Bariev's exact results demonstrate the dependence of the
local magnetisation critical exponent on the values of the defect coupling. For the ladder defect,
the local critical exponent is
\begin{equation} \label{gl:4_9}
\beta_l = \frac{2}{\pi^2} \arctan^2\left( \kappa_l^{-1} \right)
\end{equation}
with
\begin{equation} \label{gl:4_10}
\kappa_l = \tanh (\frac{J_l}{k_BT_{2d}})/\tanh (\frac{J}{k_BT_{2d}}),
\end{equation}
whereas for the chain defect one obtains
\begin{equation} \label{gl:4_11}
\beta_{ch}= \frac{2}{\pi^2} \arctan^2 \left( \kappa_{ch} \right)
\end{equation}
with
\begin{equation} \label{gl:4_12}
\kappa_{ch}= \exp \left( -2 \left( \frac{J_{ch}}{k_BT_{2d}} - \frac{J}{k_BT_{2d}} \right) \right).
\end{equation}
$J$ is the strength of the unperturbed interactions, whereas $T_{2d}$ is the critical temperature of   
the two-dimensional Ising model.
For both cases, enhanced (reduced) defect couplings yield a lower (higher) local critical
exponent as compared to the perfect two-dimensional Ising model, $\beta_{2d}=1/8$.

In the following, I discuss the influence of edges with opening angle $\theta=\pi/2$, 
see Figure \ref{fig_edge_1},
on the local critical behaviour at the surface transition. 
Because edges are one-dimensional, and all couplings
in the models are of short range, edge quantities only become singular at the temperature $T_s$
where the surface orders.
Near the surface transition, where the critical fluctuations are of two-dimensional character,
the edge then acts like a defect line in an essentially two-dimensional bulk Ising model. The edge
coupling $J_e$ corresponds to a chain-like defect, the edge-surface coupling $J_{es}$
to a ladder-type defect. The change in the topology at the edge compared to the surface amounts
to a complicated ladder-type defect.

The value of the critical exponent
is non-universal, it varies continuously as a function of the coupling strength $J_{es}$
as shown in Table \ref{table_edge_2}. These
findings may be related to the reported results for two-dimensional Ising models with a ladder
defect. The ladder corresponds to the edge, and the ladder coupling $J_l$ reflects
not only the edge-surface interaction $J_{es}$ but also the reduced connectedness to bulk spins
at the edge compared to the surfaces. For $J_{es}=J_e=J_s$ and $J_s=2 J_b$, 
the critical exponent of the edge
magnetisation has the value $\beta_2= 0.095(5)$, significantly lower than the
critical exponent of the perfect two-dimensional Ising model. Comparing the local critical exponent
$\beta_l$ near a ladder defect, see equations (\ref{gl:4_9}) and (\ref{gl:4_10}), with $\beta_2$,
one may attribute an effective ladder coupling $J_l^{eff}$ to the edge with $J_l^{eff} > J$ ($= J_s$). This
effective enhancement of the couplings is due to the influence of the bulk spins. When lowering the
coupling $J_{es}$ while keeping $J_s$ and $J_b$ constant, 
the value of $\beta$ increases, as expected from equation (\ref{gl:4_9}), see
Table \ref{table_edge_2}. 
The weakening of the edge interaction $J_e$ has a less pronounced impact on $\beta_2$ than the
weakening of $J_{es}$. The close agreement of $\beta_2$ with $\beta_{2d}$ in the case $J_e=J_{es}/2$ is
fortuitous. It is due to a compensation of a reduction in $\beta_2$ following from the increase
in the effective ladder coupling stemming from the edge
topology, and of an enhancement in $\beta_2$ following from the decrease in the strength of the
edge coupling.

\begin{table}
\begin{center}
\begin{tabular}{|c||c|c|c|}  
\hline
 & $J_{es}=J_s$, $J_e=J_s$ & $J_{es}=J_s/2$, $J_e=J_s$ & $J_e=J_s/2$, $J_{es}=J_s$\\
\hline\hline
$\beta_2$ & $0.095(5)$ & $0.176(5)$ &  $0.127(5)$\\
\hline
$J_l^{eff}$ & $1.22 \, J_s$ & $0.74 \, J_s$ & $0.99 \, J_s$\\
\hline
\end{tabular}
\end{center}
\caption
{\label{table_edge_2} Ising edge critical exponents $\beta_2$ at the surface transition for 
systems with opening angles $\theta=\pi/2$ and $J_s=2 J_b$. $J_l^{eff}$ is the effective
defect coupling of a ladder defect having the same critical exponent as the edge.
}
\end{table}

At the surface transition of three-dimensional Ising models
corner criticality deserves to be analysed
as well. As edges are local perturbations acting similar to defect lines in
two-dimensional models, the corners of a cube may be interpreted as intersection points of
three defect lines. As shown in Table \ref{table_edge_3} one again obtains critical exponents
changing continuously with the strengths of the different couplings.

\begin{table}
\begin{center}
\begin{tabular}{|c||c|c|c|}  
\hline
 & $J_{es}=J_s$, $J_e=J_s$ & $J_{es}=J_s/2$, $J_e=J_s$ & $J_e=J_s/2$, $J_{es}=J_s$\\
\hline\hline
$\beta_3$ & $0.06(1)$ & $0.26(2)$ &  $0.14(2)$\\
\hline
$\beta_{star}$ & $0.082$ & $0.210$ & $0.128$\\
\hline
\end{tabular}
\end{center}
\caption
{\label{table_edge_3} Ising corner critical exponents $\beta_3$ at the surface transition for 
systems with opening angles $\theta=\pi/2$ and $J_s=2 J_b$. 
$\beta_{star}$ is the local critical exponent near a star defect where three ladder defects
with defect couplings $J_l^{eff}$, see Table \ref{table_edge_2}, intersect.
}
\end{table}

To explain these findings, note that the corners are intersection points of edges and recall
that at the surface transition the critical fluctuations are essentially two-dimensional. The critical
exponent of the corner magnetisation, $\beta_3$, has been related to that of the magnetisation
at the intersection of three semi-infinite defect lines in the two-dimensional Ising model.
The critical exponent $\beta_{star}$
for star defects formed by three intersecting ladder defects has been calculated by Henkel {\it et al.}
\cite{Hen88b,Hen89}. One may assign to the edges an effective ladder coupling $J_l^{eff}$ by setting
$\beta_l=\beta_2$. Inserting this effective coupling into the equation derived for a star defect
formed by three ladder defects \cite{Hen88b,Hen89} then gives $\beta_{star}$. The comparison of
$\beta_3$ with $\beta_{star}$ yields a satisfactory agreement, see Table \ref{table_edge_3}.
Of course, a more refined analysis would focus on the fact that the correct geometry
of the present problem is not that of a plane but that of a cone with three defect
lines meeting at an angle $\pi/2$ \cite{Tur04} (see \cite{Cos03} for a recent discussion
of Ising models with conical singularities). Furthermore, the rather complicated nature of the edge
as a simultaneously ladder- and chain-type defect line 
as well as the effect of the bulk spins
on the corner magnetisation
should then also be taken into account.

\subsection{Wedges and surface fields}
The critical behaviour
near edges has also been studied recently at the normal transition \cite{Han99}. The authors
investigated critical adsorption near edges due to the presence of symmetry-breaking surface
fields in a wedge. Besides studying the problem in mean-field approximation, they
also presented some exact results for the two-dimensional case. Especially, the critical
edge exponent $\beta_2$ was discussed. It was shown that at the normal transition
local critical exponents also depend continuously on the opening angle.

Critical adsorption near edges is, however, only one of the many intriguing phenomena encountered
in systems with both wedges and surface fields. 
In the case of adsorption of a fluid on a solid substrate with wedge geometry \cite{Hau92,Nap92}
it is expected
from thermodynamic arguments that the liquid fills the wedge completely at temperatures
$T > T_F$, where the filling temperature $T_F$ is lower than the wetting temperature $T_W$
of a planar, but otherwise identical, wall \cite{Rej99}. In a series of recent papers Parry {\it et al.}
\cite{Par99,Par00a,Par00b,Par01a,Par01b}
studied the corner filling transition in detail, focusing especially on fluctuation effects
and on the universality classes of filling transitions. One of their prediction was that
critical filling, i.e.\ the filling transition for $T \longrightarrow T_F^-$, could be continuous
even in cases where the related wetting transition is of first order. They also studied
the divergence of various length scales associated with this phase transition and predicted,
basing themselves on a fluctuation theory, that the interfacial height $l_0$ from the bottom
of the wedge should diverge as $l_0 \sim (T_F-T)^{-1/4}$ \cite{Par00a,Bed01}.

The wedge filling transition can be studied in Ising models with wedge geometries by applying
surface fields which favor one of the two phases in the wedge 
\cite{Lip98,Abr02,Alb03,Mil03}. Exact results in the two-dimensional case \cite{Abr02}
permit to establish the existence of this kind of transition. Recent numerical investigations
\cite{Alb03,Mil03} revealed a very good agreement with the theoretical predictions 
of Parry {\it et al}. In addition, a new type of interface localisation-delocalisation
transition was revealed in the three-dimensional double wedge forming a pore with a
square cross section \cite{Mil03}. The critical exponents of this transition can
be related to the critical exponents of the filling transition in a simple wedge.

\subsection{Non-equilibrium systems}
The influence of wedges on critical behaviour has not only been studied in equilibrium
systems but also in non-equilibrium systems. Fr\"{o}jdh {\it et al.} \cite{Fro98}
introduced an edge into a directed percolation process and analysed the impact this
edge has on the non-equilibrium phase transition observed in this system.
This work generalized earlier studies of directed percolation in a semi-infinite
geometry \cite{Jan88} to the wedge geometry. Angle dependent edge critical exponents
were observed in this non-equilibrium case, too. Directed percolation processes
have also been studied in two-dimensional parabolic-like systems 
with a free surface at $y = \pm C x^k$ \cite{Kai94a,Kai95}. 
For $k< 1/z$, $z$ being the dynamical exponent,
the surface shape is a relevant perturbation and the tip order parameter displays
a stretched exponential behaviour. In the marginal case, $k=1/z$, a non-universal 
local critical behaviour is again observed.

Other non-equilibrium absorbing phase transitions have also been studied in
semi-infinite systems recently \cite{Lau98,How00}, but these studies have not
been extended to wedge-shaped geometries.

\section{Critical phenomena at non-perfect surfaces}
In the preceding Sections we discussed critical phenomena in systems
with various geometries: semi-infinite systems, wedge-shaped systems,
and cubes. All the results presented so far have in common that only
idealized, perfect surfaces were considered. 
However, real surfaces are often naturally rough, as steps or islands occur
during growth processes or result from the effect of erosion. Furthermore,
methods from nanoscience permit to create artificial structures on top of
films. Examples include lines of adatoms or regular
arrangements of geometric structures.
All these defects have some impact on
magnetic surface quantities.

In the following I review theoretical studies where surface critical phenomena
in systems with various surface defects have been investigated. Section 4.1 deals
with semi-infinite (i.e.\ bulk terminated) systems, whereas Section 4.2 is devoted to the influence of
surface defects on the critical behaviour of (ultra)thin films.

\subsection{Semi-infinite systems with surface imperfections}
Geometric surface imperfections (e.g., islands or vacancies) and impurities
may be stable on the time scale of magnetic phenomena and thus lead to
quenched surface disorder \cite{Die90a}.
Some experiments indicate an enhancement of disorder near surfaces, thus
pointing to the possible realisation of quenched surface disorder
in systems where bulk disorder is negligible.
In a theoretical description, this kind of surface disorder is usually mimicked
by random surface fields or by random surface couplings.

Early studies mainly focused on the global phase diagram observed in systems with
random surface couplings or random surface fields. In \cite{Ben85} 
dilute semi-infinite
Ising models with bond and site dilution both in the bulk and at the surface were considered.
Mean-field results obtained for semi-infinite Ising models
with random surface and bulk fields were presented in \cite{Sab87}. The possible types of phase
diagrams were established and the phase transitions as well as multicritical points
were discussed. Phase diagrams obtained for semi-infinite transverse Ising models
with random surface and bulk fields were analysed recently in \cite{Dak01}.

In a series of papers, Kaneyoshi (see, e.g., \cite{Kan88,Kan89,Kan91}) studied
semi-infinite Ising systems with an amorphous surface layer in detail. The amorphous surface 
can be mimicked
by choosing randomly weak or strong nearest-neighbour ferromagnetic couplings between surface
spins. Sometimes, he also considered random couplings between surface spins and the
underlying bulk spins. Using different analytical approaches, surface magnetic properties as, 
for example, the magnetisation of the surface layer, were analysed. The corresponding 
phase diagrams were derived from the results of an effective field theory which includes 
correlations. One interesting result obtained in
these studies concerns the location of the special transition point. Kaneyoshi observed
that the
critical coupling ratio $r_{sp}$ is shifted to higher values
in systems with a diluted surface and he conjectured that this is a common effect of surface
amorphisation. This was confirmed by a Monte Carlo study \cite{Ple98a} where
the strength of
surface bonds was chosen randomly between two different values
$J_{s1}$ and $J_{s2}$, the ratio $d =J_{s1}/J_{s2}$ measuring the degree
of dilution.
For $d=1/10$
the special transition point is located at the mean coupling ratio $r_{sp}=1.7(0.1)$
for the simple cubic lattice, noticeably
larger than the value obtained for the perfect surface $r_{sp} \approx 1.5$, see Section 2.1.
In order to understand
this shift we recall that in the limit $r=(J_{s1}+J_{s2})/(2J_b) \gg 1$ the surface effectively
decouples from the underlying bulk and can be regarded as a two-dimensional system.  However,
it is well known that at a given mean coupling $(J_{s1}+J_{s2})/2$
the critical temperature of the two-dimensional Ising model is reduced
by randomness \cite{Wan90b}.
Therefore, for fixed $r$,
increasing randomness shifts the line of the surface transition to lower temperatures which in
turn shifts the location of the special transition point to larger values of the
coupling ratio.

Sometimes, alloy surfaces are mimicked by semi-infinite surfaces with randomly
decorated magnetic and nonmagnetic atoms located on the surface \cite{Kan98}. Again,
the location of the special transition point is observed to shift as a function of the model
parameters. 
Quantities having an impact on this shift are for example the concentration of the nonmagnetic
surface atoms or the spin of the magnetic atoms.

The investigations reviewed so far almost exclusively focused on
global phase diagrams of semi-infinite systems with non-perfect surfaces. The different
authors did not try to study in detail the influence of these imperfections on the
surface critical behaviour. However,
as a certain degree of
imperfections is unavoidable when studying surfaces experimentally,
it is very important to clarify whether the
surface critical exponents are robust against surface imperfections.

A first step in this direction was taken by Mon and Nightingale in their study of the influence of
a random surface field on the surface critical behaviour of the semi-infinite Ising model \cite{Mon88}.
This work was motivated by the determination of the surface order parameter critical exponent $\beta_1$ in
a study of wetting phenomena of binary
mixtures consisting of a polar and a nonpolar liquid \cite{Dur87}. The value obtained for $\beta_1$
in that study was compatible with the value encountered in the semi-infinite Ising model with a perfect surface,
even so $\beta_1$ described the vanishing of the surface order parameter in the
presence of a random surface field with zero average strength.
Using a Harris type criterion \cite{Har74a}, Mon and Nightingale conjectured that, at the ordinary
transition, a random surface field is irrelevant for the surface critical behaviour of the Ising model.
They verified their prediction with the aid of Monte Carlo simulations. 
Using renormalization group techniques it was shown in \cite{Das88}
that surface bond-dilution is irrelevant for
the semi-infinite Ising model, the universality
classes of the different transitions being those of the pure system.

In an extensive study, Diehl and N\"{u}sser \cite{Die90a} derived Harris type criteria for
various types of quenched surface disorder with the aim to assess the relevance or irrelevance
of these random imperfections on the pure system surface critical behaviour. In his original work,
Harris \cite{Har74a} studied the stability of critical bulk systems in the presence of
randomness. The well-known Harris criterion states that, in the weak-disorder limit, bond disorder
is irrelevant for the bulk critical behaviour provided the specific heat critical exponent $\alpha$
is negative. Disorder is relevant for $\alpha > 0$, whereas the case $\alpha=0$ is marginal.
Looking at the Ising model, one concludes, based on the Harris criterion,
that in three dimensions disorder is relevant as
$\alpha \approx 0.11$, whereas in two dimensions one encounters the marginal case $\alpha=0$
which has attracted much interest \cite{Sha94}. One should note that the condition
for irrelevance of disorder resulting from a Harris type criterion is a necessary but
not a sufficient condition for the stability of the pure system's critical behaviour.
Nevertheless, a Harris type criterion is usually a reliable indicator for the irrelevance
of randomness.

The Harris criterion generalized in \cite{Die90a} to surface critical behaviour states that
short-range correlated disorder is relevant or irrelevant depending on whether
some (surface) susceptibility, which depends on the kind of randomness under investigation,
diverges or is finite at the critical point. For random surface fields the quantity of
interest is the surface susceptibility $\chi_{11}$ of the pure system, whereas for
random surface couplings the relevant susceptibility is the local specific heat $C_{11}$
\cite{Die86}. As these generalized susceptibilities have
a singularity of the form
\begin{equation}
X_{11} \sim \left| T - T_c \right|^{- \Gamma_{11}}
\end{equation}
at the critical point,
the criterion indicates that
the disorder is relevant for positive $\Gamma_{11}$ (i.e.\ $\gamma_{11}$ or $\alpha_{11}$),
but that it is irrelevant for negative $\Gamma_{11}$.

This criterion has been applied by Diehl and N\"{u}sser in various cases involving surfaces. Thus,
the presence of random surface fields is expected to be irrelevant at the ordinary transition
for all $O(n)$ models with 
$n \leq 3$ if the dimension $d > 2$, in accordance with the results obtained by Mon 
and Nightingale \cite{Mon88} for
the special case of the
semi-infinite Ising model ($n=1$). Interestingly, Feldman and Vinokur recently showed \cite{Fel02}
that weak quenched surface disorder destroys bulk ordering in the case of a system
with continuous symmetry (as the $XY$ model), leading to a power law decay of correlation
functions and therefore to the appearance of quasi-long-range order.
In the two-dimensional semi-infinite Ising model the perturbation caused by the random surface field
is marginal so that the criterion does not yield a definite answer. This case was studied subsequently
in \cite{Igl91,Car91,DeM98,Bag04} where it was shown that the 
surface critical behaviour of the 2d Ising model
is described by Ising critical exponents with logarithmic corrections to scaling.
For the special transition point, the Harris like criterion of \cite{Die90a} indicates that random
surface fields are relevant in dimensions $d \leq 4$. However, the three-dimensional semi-infinite Ising
model with quenched random surface fields does not present a special transition point nor an extraordinary
transition, as the surface
transition line, where the surface orders alone, does not exist anymore. This
follows from the fact that at finite temperatures random bulk fields destroy long-range
order in the two-dimensional bulk Ising system \cite{Nat88}.  In the cases where the extraordinary
transition still exists in the presence of random surface fields, the criterion predicts that
the disorder is then irrelevant. For random surface couplings, the perturbation is found to be irrelevant
both at the ordinary and at the extraordinary transitions. More interesting is the situation at the
special transition point where early estimates of the critical exponent $\alpha_{11}$ in three 
dimensions yielded
small negative \cite{Bin84, Eis82} {\it or} small positive values \cite{Die86}, implying that short-range
enhancement disorder is close to being relevant in three dimensions. 
Recently, new field-theoretical estimates 
also resulted in
negative values for $\alpha_{11}$ \cite{Die94, Die98a}. The Monte Carlo results obtained for
three-dimensional semi-infinite Ising models with random surface-bond disorder \cite{Ple98a} 
are also compatibel
with the irrelevance of random surface enhancement for the special transition point critical behaviour.
However, as
short-range random surface enhancement is close to being relevant, one might expect long-range correlated
enhancement disorder to be relevant at the special transition point \cite{Die90a}.

Diehl and N\"{u}sser also studied the impact of surface-enhancement disorder at the special transition
of a bulk
tricritical system. As in this case the Harris criterion did not permit a definite prediction, renormalization
group techniques were used in a second paper \cite{Die90b}
to clarify the situation.

The robustness of the surface critical exponents at the ordinary transition
against two types of surface imperfections was established in a 
Monte Carlo study \cite{Ple98a}. The studied imperfections correspond to an amorphous surface,
mimicked by random strong and weak couplings in the surface, and to a simple case of corrugation,
due to the presence of a step of monoatomic height superimposed on a perfect surface.
For the amorphous surface, the Monte Carlo results showed that the dilution, at a fixed temperature $T < T_c$,
decreases the magnetisation at and near the surface. The effective critical exponent, derived from
the magnetisation of the diluted surface, follows in the Ising case
closely that of the perfect case, yielding
the same asymptotic value $\beta_1= 0.80(1)$. Randomness in the surface couplings is therefore
irrelevant for the asymptotic behaviour of the surface magnetisation, and even of minor importance for
the corrections to scaling \cite{Ple98a}.
Similarly, the estimate for surface susceptibility critical exponent $\gamma_1$
is compatibel with the one in the perfect case. These numerical findings provide support for the conjecture
of \cite{Die90a} that surface enhancement disorder is irrelevant at the ordinary transition.
The robustness of the critical exponent $\beta_1$ against dilution observed in the numerical study
was later proven by Diehl \cite{Die98b} who derived upper and lower bounds on the surface
magnetisation of the diluted surface and showed in a rigorous way that $\beta_1$ takes
the same value in the diluted system as in the perfect system.

\begin{figure}
\centerline{
\psfig{figure=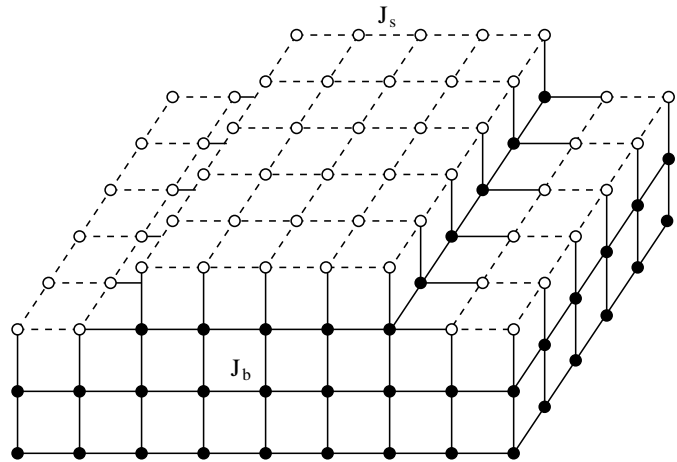,width=6cm,angle=270}}
\caption
{\label{fig_dirt_1} Geometry of a model with
two straight steps of
monoatomic height.
$J_b$ and $J_s$
are the ferromagnetic bulk and surface couplings, respectively.}
\end{figure}

In \cite{Ple98a} the effect of corrugation on surface
critical behaviour was studied in Ising systems where an extra half layer in the form of a strip-like
terrace of monoatomic height was superimposed on a perfect surface, as shown in Figure
\ref{fig_dirt_1}.
For this geometry the magnetisation at the step-edge deviates most significantly from the
magnetisation of the perfect surface. The local critical exponent describing the
behaviour of the step-edge magnetisation was found to have the value 0.800(15),
in agreement with that
of the magnetisation of the flat surface. However, in contrast to the case of random couplings,
corrections to scaling are here distinctively different from those of the perfect
surface. Diehl also considered this type of imperfections in \cite{Die98b} and showed
in a rigorous way that the critical exponent of the step-edge magnetisation is identical
to that of the magnetisation of a perfect surface.
Note that at the surface transition local critical magnetic exponents are expected
to be non-universal close to the step-edge. Indeed, the step-edge should then act like a defect line in a
system governed by two-dimensional critical fluctuations. This is supported by Monte
Carlo simulations which yield for $J_s=2J_b$
the value $0.33(2)$ for the step-edge magnetisation
critical exponent \cite{Chu00} in the Ising model.

The study of magnetic properties of rough, corrugated surfaces near criticality
is a rather new and promising field. The case of one additional terrace on an
otherwise perfect surface may be generalized to the more complex situation of
vicinal surfaces with terraces separated by equally spaced monoatomic steps 
\cite{Ben99, Zha00, Mok01,Bah02a}.
In nature, rough surfaces often result from a growth process and strongly affect
the surface magnetisation. Diffusion-limited growth results in a rough growth front
following a Poisson distribution. This may be realized in simulations by considering a surface
formed of columns of random heights \cite{Zha00}, see also \cite{Rei98a}. Layer-by-layer
growth, however, may result in films with a finite number of complete layers and one
partially filled layer,  thus yielding a different type of roughness \cite{Rei00}.
Whereas the different kinds of roughnesses discussed so far
have been shown to be (or are supposed to be) irrelevant for the surface critical
behaviour at the ordinary transition, this is not always the case. In a recent
interesting work, Hanke and Kardar \cite{Han01,Han02} showed that self-affine
rough surfaces may give rise to novel surface critical behaviour with surface
critical exponents different from those of the perfect case. Similar
results have previously been obtained in two-dimensional systems
with fractal boundaries \cite{Dup98, Car99} yielding new multifractal
boundary exponents. These results in two dimensions have been interpreted
in terms of a scale-dependent distribution of opening angles of the
fractal boundary \cite{Car99}. The same analogy has been evoked in the
discussion of the surface critical behaviour of self-affine surfaces in \cite{Han01,Han02}.

Hanke and Kardar also studied critical correlations in the vicinity of a regularly patterned surface 
(as might for example result from a lithographic preparation of the surface). Using a
perturbative calculation in the deformations in height from a flat surface they showed
that the leading power law decay of the correlations is the same as for a flat surface,
but with a modulated amplitude reflecting the shape of the surface \cite{Han01,Han02}.

Extended surface defects have been studied intensively in two dimensions.
In the Hilhorst-van Leeuwen model \cite{Hil81,Igl93} one considers a semi-infinite
square lattice with inhomogenous couplings (alternatively, the triangular
lattice has also been investigated). In the direction parallel to the surface
one has couplings with a constant strength $J_1$, whereas the strength of the
couplings varies perpendicular to the surface as a function of the distance $y$
to the surface:
\begin{equation} \label{eq_hvl}
J_2(y) - J_2(\infty) = \frac{A}{y^\omega}.
\end{equation}
This extended perturbation is irrelevant for $\omega > 1/\nu_b =1$, yielding
the same critical behaviour as the homogeneous semi-infinite system. For $\omega < 1$,
however, the perturbation is relevant and a spontaneous surface magnetisation
is observed at the bulk critical point. The most interesting case is $\omega =1$,
where the perturbation is marginal. Here a spontaneous surface magnetisation is
again observed for values of $A$ larger than some threshold $A_c$, whereas for
$A < A_c$ the scaling dimension $x_1$ varies continuously as a function
of $A$: $x_1 = \frac{1}{2} ( 1- A/A_c)$. This intriguing behaviour has attracted
much interest in the past \cite{Igl93} and has recently led to the discovery
of an up to then unnoticed effect in the short-time non-equilibrium critical dynamics
at surfaces \cite{Ple03}.

Finally, before turning to thin films with surface imperfections,
let us briefly mention some recent work concerning surface critical behaviour
in an Ising model with quenched random defects in the bulk. In the three-dimensional
bulk Ising model, quenched random defects are a relevant perturbation leading
to modified critical bulk exponents \cite{Har74b,Gri76}. The surface critical behaviour
of the three-dimensional Ising model with quenched bulk disorder has been investigated
both at the ordinary \cite{Ohn92,Usa01} and at the special transition point \cite{Ohn92,Usa02},
using different renormalization group techniques. The crossover behaviour between these two 
transitions was studied recently in \cite{Usa03}. Interestingly, modified surface critical
behaviour is encountered independently whether quenched surface disorder is present or not.
The corresponding marginal case $d=2$ has also been studied in detail
recently \cite{Igl98, Laj00}, yielding Ising surface critical exponents
with logarithmic corrections to scaling. Boundary critical behaviour of $q$-state
random Potts models hase been studied in \cite{Pal00} for $3 \leq q \leq 8$.
It should be noted that the problem of bond percolation has also been investigated
in semi-infinite geometries \cite{The79,DeB80,
Car80,Chr86,Die89}. In this case distinct surface percolation transitions have
been identified and the values of the local critical exponents have been computed.

\subsection{Thin films with surface imperfections}
In a series of papers Aar\~{a}o Reis studied
the dependence of physical quantities in two-dimensional Ising stripes \cite{Rei97, Rei98b,
Rei98c, Rei01} and three-dimensional Ising
thin films \cite{Rei98a, Rei00} on the surface roughness.
Using transfer matrix methods, Reis investigated stripes where the column
heights were chosen according to a Gaussian distribution with mean $L$ ($L$: integer) and
variance $(\Delta L )^2/2$ \cite{Rei97, Rei98b}. Two different cases were considered:
$\Delta L$ constant and $\Delta L = L/L_0$ with some constant $L_0$. Whereas the roughness becomes
unimportant for large $L$ in the former case, in the latter case the roughness increases
with the mean thickness. Reis paid special attention to the finite-size
scaling and to the finite-size corrections due to randomness. In \cite{Rei98c} he also
considered the case of noninteger mean $L$ and showed that the free energy displays an
interesting oscillating behaviour as function of continuously changing $L$ due to
oscillations in the mean coordination number. Recently \cite{Rei01}, he completed
his investigation of stripes of random width by studying roughness of the more
general form $\Delta L \sim L^\beta$ with $0 \leq \beta \leq 1$.
He also considered correlated roughness by imposing the maximal
height difference between neighbouring columns to be not larger than one lattice constant.
Interestingly, these computations showed that the correlations had no systematic
effect on the corrections to scaling. More relevant to the understanding of
the critical behaviour of real films
are Reis' studies of films with rough surfaces \cite{Rei98a,Rei00}.
Using Monte Carlo techniques, he studied the same kind of uncorrelated roughness as
for the stripes: Gaussian distribution of thicknesses with integer mean $L$ and
$\Delta L$ constant for all $L$ or $\Delta L = L/L_0$ \cite{Rei98a} as well as
noninteger mean $L$ \cite{Rei00}. These studies were partly motivated by the observation
that some film quantities depend on the growth condition, the different growth mechanism
leading to different types of roughness. Reis studied the magnetic susceptibility,
the specific heat, and the total magnetisation of films with different thicknesses.
Especially, he showed that the simple equation (\ref{gl:5_1}) connecting the critical
temperature of finite films with the bulk critical temperature does not hold for
noninteger mean thicknesses $L$, as $T_c(L)$ shows a convex behaviour between two
consecutive integer values of $L$ \cite{Rei00}. He also studied the critical behaviour
of rough films and thereby observed  that the considered types of
disorder are irrelevant: as for flat films, the critical
exponents of the two-dimensional Ising model are recovered in rough thin films.
Up to now only rough films with uncorrelated roughness have been
studied. It would be very interesting to investigate the influence of spatial correlated roughness
on the critical thin film behaviour as well.

The impact of additional regular structures, located on the surface, on the critical behaviour
of Ising films is studied numerically in \cite{Chu00,Sel02a}.
The additional structures consist of one or two adjacent lines formed by adatoms as well as
straight steps of unit height.
One may introduce local couplings with different strengths, similar to what has been done in Section 3.1 for
the edges. For example, the strength of the bonds connecting two neighbouring
adatoms in the defect line or the strength of the coupling between the adatoms and the
underlying magnetic film may be varied. 
The main finding of \cite{Chu00} is that non-universal critical behaviour is 
encountered in Ising
films with additional surface defects: the local magnetic critical exponents close to the defects
depend continuously on the local couplings as well as on the layer thickness.
Interestingly, the presence of an additional line is already enough to change the local critical 
exponent of a one-layer system
by a large amount. Indeed, in the case where all the couplings have equal strength
one obtains the value $\beta_l \approx 0.084 < 1/8 = \beta_{2d}$ for the critical exponent
of the local order parameter \cite{Chu00}. 

It is worthwhile mentioning
that related experiments on films with additional adatom lines have been published \cite{She97,She03,Sko03},
even if there the lines of adatoms were not connected magnetically to the film.
The magnetism of lines of $4d$ adatoms on Ag surfaces
has been the subject of some theoretical studies \cite{Baz00,Bel01}. The magnetic behaviour
of thin films with large terraces has also been analysed experimentally \cite{And01}.

Increasing the film thickness further, the semi-infinite system with surface imperfections
is reached at the end. The following two typical scenarios must be distinguished: (i) for
$J_s/J_b < r_{sp}$, bulk and surface order at the same temperature
(ordinary transition), whereas (ii) for $J_s/J_b > r_{sp}$ the surface orders at a higher
temperature than the bulk (surface transition). $r_{sp}$ is the critical coupling ratio
of the multicritical point, see Section 2.1. Therefore,
a completely different behaviour of local quantities near the surface imperfections
is expected in both cases when varying the number of layers $L$.
Numerical simulations of non-perfect films are in complete agreement with this expectation \cite{Chu00}.
For $J_s/J_b < r_{sp}$ the surface critical exponent $\beta_1 \approx 0.80$
is obtained in the limit $L \longrightarrow \infty$ everywhere on the surface, independently
of any additional surface structure, as discussed already in Section 4.1. In a film
the local magnetisation near the defect closely follows the local magnetisation of the corresponding
semi-infinite system for low temperatures. At temperatures where the correlation length is comparable
with the thickness of the film a crossover from a regime
with isotropic three-dimensional fluctuations to a regime
dominated by two-dimensional critical fluctuations takes place.
Increasing the film thickness the crossover temperatures approaches the bulk critical
temperature, yielding in the limit of the
semi-infinite system, $L \longrightarrow \infty$, the value $\beta_l=\beta_1 \approx 0.80$
for the critical exponent of the local magnetisation.
Choosing $J_s/J_b > r_{sp}$, non-universal local critical behaviour close to the adchain is
expected at the surface transition even in the limit $L \longrightarrow \infty$.
This situation is indeed
comparable to that of an edge at the surface transition, the edge corresponding to an
extended defect line as discussed in Section 3.2. The local magnetic critical exponents
should therefore depend continuously on the local couplings at the additional line.
This has been studied in \cite{Chu00} by examining the case $J_s = 2 J_b$.   
The results obtained show that near an additional line $\beta_l$
is affected both by the values of the local couplings and by the presence of additional
bulk-like layers leading to non-universal local critical behaviour.

\begin{figure}
\centerline{
\psfig{figure=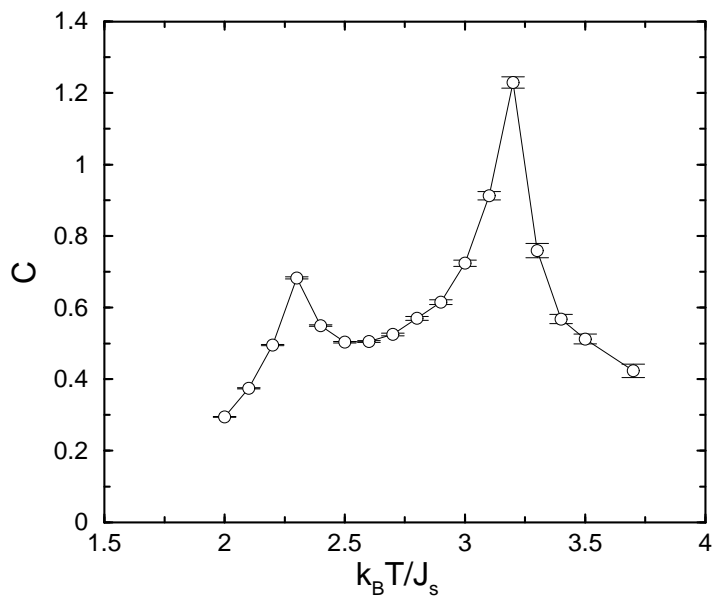,width=8cm,angle=270}}
\caption
{\label{fig_dirt_2}
Specific heat of an Ising film with equal couplings consisting of one layer plus half a layer.
Two distinct peaks are observed corresponding to two distinct phase transitions taking place in this
composite system.}
\end{figure}

Further results have been obtained for systems with a straight step on top of film.
At the ordinary transition, a straight step on top of a semi-infinite system does not change
the local critical exponents \cite{Ple98a}. In thin films the critical behaviour is however quite
different \cite{Chu00, Sel02a}. Introducing a straight step 
by adding half a layer of magnetic adatoms
to the surface of the magnetic film one observes two sharp peaks
in the specific heat \cite{Chu00, Sel02a,And01}, see Figure \ref{fig_dirt_2},
pointing to the existence of two
different phase transitions. In fact, when considering a system with 
one half layer on top of a film 
one is dealing with a composite system displaying in the thermodynamic limit two distinct phase
transitions at two different temperatures. Consider for simplicity the case of a 
single layer plus half a layer. One phase transition then takes place 
at the critical temperature of the 2d Ising model, $k_BT_c(L=1)/J_s=2.269...$ and one
at the critical temperature of the double layer, $k_BT_c(L=2)/J_s=3.207(3)$. The value
of the critical exponent of the step magnetisation at the higher temperature phase transition
is 1/2, i.e.\ is identical to the value of the surface critical exponent of the
two-dimensional Ising model. At the lower critical
temperature the single semi-infinite layer then orders in presence of ordered surface spins, the
ordered double layer acting at that transition as a surface field. One is therefore dealing
with the normal transition. The same scenario is expected to hold for finite films with an
additional half layer with $T_c(L+1) > T_c(L)$, yielding at $T_c(L+1)$ the critical exponent
1/2 near the step edge, in accordance with the numerical findings of \cite{Chu00} and \cite{Sel02a}.
Note that very recently \cite{And01}, two sharp susceptibilities peaks observed 
experimentally in thin
films with large terraces have been interpreted as the signatures of two distinct phase transitions. 

Thin films with an additional terrace covering half of the surface are only one
example of composite systems displaying two distinct phase transitions. Further
examples include two-dimensional Ising models with a defect column where the bonds
differ on the two sides of the column \cite{Bar90}, Ising quantum chains where
couplings and transverse fields differ in the two half chains \cite{Ber91}, or layered
magnetic systems consisting of thin layers of coupled Ising spins with $S = 1/2$ and
$S=1$ \cite{Kar97}. In \cite{Igl90}, wetting phenomena were studied in models consisting
of two semi-infinite systems with unequal critical temperatures connected by a defect plane.

Finally, let us briefly mention that 
in the context of spin systems with continuous spin symmetry
thin films with selected surface defects
(amorphisation of the surface layer \cite{ Ham03} or steps \cite{Zha02}) have been
considered in some cases, but no systematic
study of the critical behaviour of this kind of systems has been done up to now.
 
\section{Concluding remarks}
Surface criticality has been the subject of intensive studies in the last
thirty years. Thereby a large variety of different methods (analytical,
numerical, and experimental) has been used, yielding a host of interesting
results, as reviewed in this work. 
Critical phenomena at perfect surfaces are now in general well
understood, at least when dealing with static critical quantities. This is
not really the case for dynamic critical behaviour at surfaces for which a
coherent picture has not yet emerged. A major problem in this context is the
total lack of experimental studies on surface dynamic properties at criticality.
The situation is also not very satisfactory from the theoretical point of view, 
as only selected results on equilibrium and non-equilibrium surface dynamic
behaviour at criticality have been published. Clearly, this is one of the
most important aspects of surface criticality that warrants more attention 
in the future.

Wedge-shaped geometries, which can be viewed as generalisations of semi-infinite
systems, have also been discussed in detail in this review. It is encouraging
that the effect of curved surfaces on the local critical behaviour has been
observed in simulations of liquid-vapour transitions near a weakly attractive surface. This
may point to possible experimental systems where this kind of problems can be
studied.
There are indeed a vast number of theoretical predictions for this kind of geometry,
and experimental investigations are therefore welcomed.

In the last years the focus of research on surface criticality has somehow shifted,
as the main emphasis has been on more realistic surfaces. The facts that
real surfaces are usually rough, displaying a variety of different surface defects,
and that experimental physicists can create artificial structures on top of 
a surface directly lead to the question whether these quantities have an impact
on local critical behaviour. We have presented a comprehensive overview of the field,
thereby discussing in detail critical behaviour in semi-infinite systems with
surface defects as well as in thin films with additional surface structures.
Some common surface defects have been shown to be irrelevant for the surface critical
behaviour at the ordinary transition. There are however some interesting exceptions, as
for example the case of self-affine rough surfaces. The situation is even more complex
at the surface transition where in a three-dimensional system the critical 
fluctuations are of two-dimensional nature. Indeed, additional structures as steps
or lines of adatoms have been shown in numerical studies to lead to non-universal
local critical behaviour where the values of the local critical exponents reflect
the strengths of the coupling constants  as well as the presence of the disordered bulk.
It is worth noting that in thin films this kind of additional surface structures
in general leads to non-universal critical behaviour.

It is obvious from our overview that a large number of recent studies of the effects
of surface defects on the local critical behaviour are either of purely numerical
nature or are using rather crude approximations. This is especially the case when 
dealing with non-perfect surfaces at the surface transition. There is a need
for more elaborate analytical approaches, and it is one of the intention of this review to
encourage further theoretical (and also experimental) investigations of critical
phenomena at non-perfect surfaces.

\ack
It is a pleasure to thank all my collaborators who worked with me on various
aspects of surface criticality: F.\ \'{A}.\ Bagam\'{e}ry,
D. Catrein, M.-C. Chung, F.\ Igl\'{o}i, M. Kaulke, I. Peschel, W. Selke, and L. Turban. 
I am indebted to W. Selke for introducing me to the field of surface critical
behaviour and for many years of fruitful collaboration. I also thank H.W. Diehl,
M.\ Henkel, and A.\ H\"{u}ller for many inspiring discussions.


\section*{References}
\begin{thebibliography}{999}
\bibitem{Bin83} Binder K 1983
{\sl Phase Transitions and Critical Phenomena} vol~8 
(London/New York: Academic Press) p.~1
\bibitem{Abr86} Abraham A B 1986
{\sl Phase Transitions and Critical Phenomena} vol~10
(London/New York: Academic Press) p.~1
\bibitem{Die86} Diehl H W 1986
{\sl Phase Transitions and Critical Phenomena} vol~10
(London/New York: Academic Press) p.~75
\bibitem{Die97} Diehl H W 1997 {\it Int.\ J.\ Mod.\ Phys.} B {\bf 11} 3503 
\bibitem{Dos92} Dosch H 1992 {\it Critical Phenomena at Surfaces and Interfaces}
(Berlin, Heidelberg, and New York: Springer)
\bibitem{Fis77} Fisher M E and Caginalp G 1977 {\it Commun.\ Math.\ Phys.}
{\bf 56} 11
\bibitem{Cag79} Caginalp G and Fisher M E 1979 {\it Commun.\ Math.\ Phys.}
{\bf 65} 247
\bibitem{Ple98a} Pleimling M and Selke W 1998 {\it Eur.\ Phys.\ J.} B {\bf 1} 385
\bibitem{Lub75} Lubensky T C and Rubin M H 1975 \PR B {\bf 12} 3885
\bibitem{Nak82} Nakanishi H and Fisher M E 1982 \PRL {\bf 49} 1565
\bibitem{Mil71} Mills D L 1971 \PR B {\bf 3} 3887
\bibitem{Bin72} Binder K and Hohenberg P C 1972 \PR B {\bf 6} 3461
\bibitem{Bin74} Binder K and Hohenberg P C 1974 \PR B {\bf 9} 2194
\bibitem{Bur77} Burkhardt T W and Eisenriegler E 1977 \PR B {\bf 16} 3213
\bibitem{Lip81} Lipowsky R and Wagner H 1981 {\it Z.\ Phys.} B {\bf 42} 355
\bibitem{Bin84} Binder K and Landau D P 1984 \PRL {\bf 52} 318
\bibitem{Rug92} Ruge C, Dunkelmann A and Wagner F 1992 \PRL {\bf 69} 2465
\bibitem{Rug93} Ruge C, Dunkelmann A, Wagner F and Wulf J 1993
{\it J.\ Stat.\ Phys.} {\bf 73} 293
\bibitem{AuY73} Au-Yang H 1973 {\it J.\ Math.\ Phys.} {\bf 14} 937
\bibitem{Mer66} Mermin N D and Wagner H 1966 \PRL {\bf 17} 1133
\bibitem{Kos73} Kosterlitz J M and Thouless D J 1973 \JPC {\bf 6} 1181
\bibitem{Kos74} Kosterlitz J M 1974 \JPC {\bf 7} 1046
\bibitem{Fro86} Fr\"{o}hlich J and Pfister C E 1986 {\it Commun.\ Math.\ Phys.}
{\bf 107} 337
\bibitem{Lan89} Landau D P, Panday R and Binder K 1989 \PR B {\bf 39} 12302
\bibitem{Pec91} Peczak O and Landau D P 1991 \PR B {\bf 43} 1048
\bibitem{Die84} Diehl H W and Eisenriegler E 1984 \PR B {\bf 30} 300
\bibitem{Aff00} Affleck I 2000 \JPA {\bf 33} 6473
\bibitem{Bak01} Bakchich A and El Bouziani M 2001 \PR B {\bf 63} 064408 
\bibitem{Igl99} Igl\'{o}i F and Carlon E 1999 \PR B {\bf 59} 3783
\bibitem{Ben97} Benayad N and Dakhama A 1997 \PR B {\bf 55} 12276
\bibitem{Kar97} Karevski D and Henkel M 1997 \PR B {\bf 55} 6429
\bibitem{The79} Theumann A 1979 \PR B {\bf 19} 6295
\bibitem{DeB80} De'Bell K and Essam J W 1980 \JPC {\bf 13} 4811 
\bibitem{Car80} Carton J P 1980 {\it J.\ Phys.\ (Paris)} {\bf 41}, L175
\bibitem{Chr86} Christou A and Stinchcombe R B 1986 \JPA {\bf 19} 757
\bibitem{Die89} Diehl H W and Lam P M 1989 {\it Z.\ Phys.} B {\bf 74} 395
\bibitem{Lip82} Lipowsky R 1982 \PRL {\bf 49} 1575
\bibitem{Laj81} Lajzerowicz J 1981 {\it Ferroelectrics} {\bf 35} 219
\bibitem{Tur02} Turban L and Igl\'{o}i F 2002 \PR B {\bf 66} 014440 
\bibitem{Wat00} Watson G M, Gibbs D, Lander G H, Gaulin B D, Berman L E, Matzke H
and Ellis W 2000 \PR B {\bf 61} 8966
\bibitem{Lip82b} Lipowsky R 1982 {\it Z.\ Phys.} B {\bf 45} 229
\bibitem{Dob04} Dobrovolny C, Laanait L and Ruiz J 2004 {\it J.\ Stat.\ Phys.} {\bf 114} 1269
\bibitem{Die94} Diehl H W and Shpot M 1994 \PRL {\bf 73} 3431
\bibitem{Die98a} Diehl H W and Shpot M 1998 \NP {\bf B528} 595
\bibitem{DeB90} De'Bell K, Lookman T and Whittington S G 1990 \PR A {\bf 41} 682
\bibitem{Heg94} Hegger R and Grassberger P 1994 \JPA {\bf 27} 4069
\bibitem{Lan90} Landau D P and Binder K 1990 \PR B {\bf 41} 4633
\bibitem{Rug95} Ruge C and Wagner F 1995 \PR B {\bf 52} 4209
\bibitem{Nig88} Nightingale M P and Bl\"{o}te H W J 1988 \PRL {\bf 60} 1562
\bibitem{Kre00} Krech M 2000 \PR B {\bf 62} 6360
\bibitem{Ber03} Berche B 2003 \JPA {\bf 36} 585
\bibitem{Sig86} Sigl L and Fenzl W 1986 \PRL {\bf 57} 2191
\bibitem{Mai90} Mail\"{a}nder L, Dosch H, Peisl J and Johnson R L 1990 \PRL {\bf 64} 2527
\bibitem{Bur93} Burandt B, Press W and Hauss\"{u}hl S 1993 \PRL {\bf 71} 1188
\bibitem{Alv82} Alvarado S, Campagna M and Hopster H 1982 \PRL {\bf 48} 51
\bibitem{Die81a} Diehl H W and Dietrich S 1981 {\it Z.\ Phys.} B {\bf 42} 65 
\bibitem{Car84a} Cardy J 1984 \NP {\bf B240 [FS12]} 514
\bibitem{Res00} Re\u{s} I and Straley J P 2000 \PR B {\bf 61} 14425
\bibitem{Den03} Deng Y and Bl\"{o}te H W J 2003 \PR E {\bf 67} 066116
\bibitem{Die81b} Diehl H W and Dietrich S 1981 \PR B {\bf 24} 2878
\bibitem{Rau87} Rau C and Robert M 1987 \PRL {\bf 58} 2714
\bibitem{Tan93} Tang H, Weller D, Walker T G, Scott J C, Chappert C,
Hopster H, Pang A W, Dessau D S and Pappas D P 1993 \PRL {\bf 71} 444
\bibitem{Tob98} Tober E D, Palomares F J, Ynzunza R X, Denecke R,
Morais J, Wang Z, Bino G, Liesegang J, Hussain Z and Fadley C S 1998 \PRL
{\bf 81} 2360 
\bibitem{Shi00} Shick A B, Pickett W E and Fadley C S 2000 \PR B {\bf 61} R9213
\bibitem{Rau83} Rau C 1983 {\it J.\ Magn.\ Magn.\ Mater.} {\bf 31-34} 874
\bibitem{Rau88} Rau C, Jin C, and Roberts M 1988 {\it J.\ Appl.\ Phys.} {\bf 63} 3667
\bibitem{Mam87} Mamaev Y A, Petrov U V and Starovoitov S A 1987
{\it Sov.\ Phys.\ Tech.\ Phys.\ Lett.} {\bf 13} 642
\bibitem{Mar99} Marynowski M, Franzen W, El-Batanouny M and
Staemmler V 1999 \PR B {\bf 60} 6053
\bibitem{El02} El-Batanouny M 2002 \JPCM {\bf 14} 6281
\bibitem{Mur03} Murphy B M, Stettner J, Traving M, Sprung M, Grotkopp I, M\"{u}ller M,
Oglesby C S, Tolan M and Press W 2003 {\it Physica} B {\bf 336} 103
\bibitem{Pol95} Polak M, Rubinovich L and Deng J 1995 \PRL {\bf 74} 4059
\bibitem{Arn00} Arnold C S and Pappas D P 2000 \PRL {\bf 85} 5202 
\bibitem{Bra77} Bray A J and Moore M A 1977 \JPA {\bf 10} 1927
\bibitem{Bur94} Burkhardt T W and Diehl H W 1994 \PR B {\bf 50} 3894
\bibitem{Dre97} Drewitz A, Leidl R, Burkhardt T W and Diehl H W 1997
\PRL {\bf 78} 1090 
\bibitem{Abr00} Abraham D B and Upton P J 2000 \PRL {\bf 85} 2541
\bibitem{Pou99} Poulopoulos P and Baberschke K 1999 \JPCM {\bf 11} 9495
\bibitem{Ou97} Ou J T, Wang F and Lin D L 1997 \PR B {\bf 56} 2805
\bibitem{Hen98} Henkel M, Andrieu S, Bauer P and Piecuch M 1998 \PRL {\bf 80} 4783
\bibitem{Sab00} Saber A, Ainane A, Dujardin F, El Aouad N, Saber M and
St\'{e}b\'{e} B 2000 \JPCM {\bf 12} 43
\bibitem{Zha01} Zhang R and Wills R F 2001 \PRL {\bf 86} 2665
\bibitem{Cab02} Cabral Neta J, Ricardo de Sousa J and Plascak J A 2002
\PR B {\bf 66} 064417
\bibitem{Bar83} Barber M N 1983
{\sl Phase Transitions and Critical Phenomena} vol~8
(London/New York: Academic Press) p.~145
\bibitem{Kre91} Krech M and Dietrich S 1991 \PRL {\bf 66} 345
\bibitem{Kre92a} Krech M and Dietrich S 1992 \PR A {\bf 46} 1886
\bibitem{Kre92b} Krech M and Dietrich S 1992 \PR A {\bf 46} 1922
\bibitem{Li92} Li Y and Baberschke K 1992 \PRL {\bf 68} 1208
\bibitem{Nic00} Nickel B, Donner W, Dosch H, Detlefs C and Gr\"{u}bel G
2000 \PRL {\bf 85} 134
\bibitem{Sch96} Schilbe P, Siebentritt S and Rieder K.-H. 1996
{\it Phys. Lett.} A {\bf 216} 20
\bibitem{Chu00} Chung M-C, Kaulke M, Peschel I, Pleimling M, and Selke W 2000
{\it Eur.\ Phys.\ J.} B {\bf 18} 655
\bibitem{Mar00} Marqu\'{e}s M I and Gonzalo J A 2000 {\it Eur.\ Phys.\ J.} B {\bf 14} 317
\bibitem{Mou03} Moussa N and Bekhechi S 2003 {\it Physica} A {\bf 320} 435
\bibitem{Sch94a} Schulz B and Baberschke K 1994 \PR B {\bf 50} 13467
\bibitem{Guo00} Guo W, Shi L P and Lin D L \PR B {\bf 62} 14259
\bibitem{Hal98}
Halilov S V, Perlov A Y, Oppeneer P M, Yaresko A N and Antonov V N 1998
\PR B {\bf 57} 9557
\bibitem{Qiu94} Qiu Z Q, Pearson J and Bader S D 1994 \PR B {\bf 49} 8797
\bibitem{Hua94} Huang F, Kief M T, Mankey G J and Willis R F 1994
\PR B {\bf 49} 3962
\bibitem{Pop01} Popov A P and Pappas D P 2001 \PR B {\bf 64} 184401
\bibitem{Bak82} Bak P 1982 {\it Rep.\ Prog.\ Phys.} {\bf 45} 587
\bibitem{Sel88} Selke W 1988 {\it Phys.\ Rep.} {\bf 170} 213
\bibitem{Yeo88} Yeomans J M 1988 {\it Solid State Physics} {\bf 41} 151
\bibitem{Sel92} Selke W 1992 {\sl Phase Transitions and Critical Phenomena} vol~15
(London/New York: Academic Press) p.~1
\bibitem{Neu98} Neubert B, Pleimling M and Siems R 1998 {\it Ferroelectrics}
{\bf 208-209} 141
\bibitem{Hor75a} Hornreich R M, Luban M and Shtrikman S 1975 \PRL {\bf 35} 1678
\bibitem{Ell61} Elliott R J 1961 \PR {\bf 124} 346
\bibitem{Fis80} Fisher M E and Selke W 1980 \PRL {\bf 44} 1502
\bibitem{Sel84} Selke W and Duxbury P M 1984 {\it Z.\ Phys.} B {\bf 57} 49
\bibitem{Sel00} Selke W, Catrein D and Pleimling M 2000 \JPA {\bf 33} L459
\bibitem{Sel02a} Selke W, Pleimling M, Peschel I, Kaulke M, Chung M-C
and Catrein D 2002 {\it J.\ Magn.\ Magn.\ Mater.} {\bf 240} 349
\bibitem{Sel02b} Selke W, Pleimling M and Catrein D 2002 {\it Eur.\ Phys.\ J.} B {\bf 27} 321
\bibitem{Mel03} Mello V D, Chianca C V, Dantas A L and Carri\c{c}o A S 2003 \PR
B {\bf 67} 012401 
\bibitem{Cha03} Charnaja E V, Ktitorov S A and Pogorelova O S 2003 {\it cond-mat/0303294}
\bibitem{Die02a} Diehl H W 2002 {\it Acta Physica Slovaca} {\bf 52} 271
\bibitem{Gum86} Gumbs G 1986 \PR B {\bf 33} 6500
\bibitem{Bin99} Binder K and Frisch H L 1999 {\it Eur.\ Phys.\ J.} B {\bf 10} 71
\bibitem{Fri00} Frisch H L, Kimball J C and Binder K 2000 \JPCM {\bf 12} 29
\bibitem{Jac01} Jacobs A E, Mukamel D and Allender D W 2001 \PR E {\bf 63} 021704 
\bibitem{Ple02b} Pleimling M 2002 \PR B {\bf 65} 184406
\bibitem{Die03a} Diehl H W, Rutkevich S and Gerwinski A 2003 \JPA {\bf 36} L243
\bibitem{Die03b} Diehl H W, Gerwinski A and Rutkevich S 2003 \PR B {\bf 68} 224428
\bibitem{Gar76} Garel T and Pfeuty P 1976 \JPC {\bf 9} L245
\bibitem{Kum76} Kumar P and Maki K 1976 \PR B {\bf 13} 2011
\bibitem{Die83} Dietrich S and Diehl H W 1983 {\it Z. Phys.} B {\bf 51} 343
\bibitem{Kik85} Kikuchi M and Okabe Y 1985 \PRL {\bf 55} 1220
\bibitem{Rie85} Riecke H, Dietrich S and Wagner H 1985 \PRL {\bf 55} 3010
\bibitem{Xio89a} Xiong G M and Gong C D 1989 \JPCM {\bf 1} 8673
\bibitem{Xio89b} Xiong G M and Gong C D 1989 {\it Z. Phys.} B {\bf 74} 379
\bibitem{Fra89} Frank D and Dohm V 1989 \PRL {\bf 62} 1864
\bibitem{Bin91} Binder K and Frisch H L 1991 {\it Z. Phys.} B {\bf 84} 403
\bibitem{Die92} Diehl H W and Janssen H K 1992 \PR A {\bf 45} 7145 
\bibitem{Die94a} Diehl H W 1994 \PR B {\bf 49} 2846
\bibitem{Wic95} Wichmann F and Diehl H W 1995 {\it Z. Phys.} B {\bf 97} 251
\bibitem{Rit95} Ritschel U and Czerner P 1995 \PRL {\bf 75} 3882
\bibitem{Maj96} Majumdar S N and Sengupta A M 1996 \PRL {\bf 76} 2394
\bibitem{Muk99} Mukherji S 1999 {\it Eur.\ Phys.\ J.} B {\bf 8} 423
\bibitem{Kre01} Krech M, Karl H and Diehl H W 2001 {\it Physica} A {\bf 297} 64
\bibitem{Die02} Diehl H W, Krech M and Karl H 2002 \PR B {\bf 66} 024408
\bibitem{Ple03} Pleimling M and Igl\'{o}i F 2004 \PRL {\it in press} [{\it cond-mat/0312583}]
\bibitem{Hoh77} Hohenberg P C and Halperin B I 1977 {\it Rev.\ Mod.\ Phys.}
{\bf 49} 435
\bibitem{Jan89} Janssen H K, Schaub B, Schmittmann B 1989
{\it Z. Phys.} B {\bf 73} 539
\bibitem{Die85} Dietrich S and Wagner H 1985 {\it Z. Phys.} B {\bf 59} 35
\bibitem{Die90} Dietrich S 1990 {\sl Magnetic properties of low-dimensional systems II}
(Berlin/Heidelberg: Springer) p.~150
\bibitem{Die88} Dietrich S 1988
{\sl Phase Transitions and Critical Phenomena} vol~12
(London/New York: Academic Press) p.~1
\bibitem{Bon01} Bonn D and Ross D 2001 {\it Rep.\ Prog.\ Phys.} {\bf 64} 1085
\bibitem{Bin03} Binder K, Landau D and M\"{u}ller M 2003 {\it J.\ Stat.\ Phys.} {\bf 110} 1411
\bibitem{Sym81} Symanzik K 1981 \NP {\bf B190 [FS 3]} 1
\bibitem{McA93} McAvity D M and Osborn H 1993 \NP {\bf B406 [FS]} 655
\bibitem{Rit96} Ritschel U and Czerner P 1996 \PRL {\bf 77} 3645
\bibitem{Cze97a} Czerner P and Ritschel U 1997 {\it Int.\ J.\ Mod.\ Phys.} B {\bf 11} 2075
\bibitem{Cze97b} Czerner P and Ritschel U 1997 {\it Physica} A {\bf 237} 240
\bibitem{Cia97} Ciach A and Ritschel U 1997 \NP {\bf B489} 653 
\bibitem{Mac99} Macio{l}ek A, Ciach A and Drzewi\'{n}ski 1999 \PR E {\bf 60} 2887
\bibitem{Mac03} Macio{l}ek A, Evans R and Wilding N B 2003 {\it J.\ Chem.\ Phys.}
{\bf 119} 8663
\bibitem{Cho01} Cho J-H J and Law B M 2001 \PRL {\bf 86} 2070
\bibitem{Bre83} Br\'{e}zin E and Leibler S 1983 \PR B {\bf 27} 594
\bibitem{Rit98} Ritschel U \PR {\bf B} {\bf 57} R693
\bibitem{Kri97} Krimmel S, Donner W, Nickel B, Dosch H, 
Sutter C and Gr\"{u}bel G 1997 \PRL {\bf 78} 3880 
\bibitem{Sch93} Schmid F 1993 {\it Z.\ Phys.} B {\bf 91} 77 
\bibitem{Lei98} Leidl R and Diehl H W 1998 \PR B {\bf 57} 1908
\bibitem{Car83} Cardy J 1983 \JPA {\bf 16} 3617
\bibitem{Igl93} Igl\'{o}i F, Peschel I and Turban L 1993 {\it Adv.\ Phys.} {\bf 42} 683
\bibitem{Bar84} Barber M N, Peschel I and Pearce P A 1984 {\it J.\ Stat.\ Phys.}
{\bf 37} 497
\bibitem{Pes85} Peschel I 1985 \PL {\bf 110A} 313
\bibitem{Kai89} Kaiser C and Peschel I 1989 {\it J.\ Stat.\ Phys.} {\bf 54} 567
\bibitem{Dav91} Davies B and Peschel I 1991 \JPA {\bf 24} 1293
\bibitem{Dav97} Davies B and Peschel I 1998 {\it Ann.\ Physik} {\bf 6} 187
\bibitem{Abr94} Abraham D B and Latr\'{e}moli\`{e}re F 1994 \PR E {\bf 50} R9
\bibitem{Abr95} Abraham D B and Latr\'{e}moli\`{e}re F 1995 {\it J.\ Stat.\ Phys.}
{\bf 81} 539
\bibitem{Abr96} Abraham D B and Latr\'{e}moli\`{e}re F 1996 \PRL {\bf 76} 4813
\bibitem{Gut84} Guttmann A J and Torrie G M 1984 \JPA {\bf 17} 3539
\bibitem{Dup86} Duplantier B and Saleur H 1986 \PRL {\bf 57} 3179
\bibitem{Kar97a} Karevski D, Lajk\'{o} P and Turban L 1997 {\it J.\ Stat.\ Phys.}
{\bf 86} 1153
\bibitem{Pes91} Peschel I, Turban L and Igl\'{o}i F 1991 \JPA {\bf 24} L1229
\bibitem{Lan76} Landau D P 1976 \PR B {\bf 14} 255
\bibitem{Mer91} Merikoski J, Timonen J, Manninen M and Jena P 1991 \PRL
{\bf 66} 938
\bibitem{Agu95} Aguilera-Granja F, Mor\'{a}n-L\'{o}pez J L and Montejano-Carrizales J M
1995 {\it Surf.\ Sci.} {\bf 326} 150
\bibitem{Sax87a} Saxena V K 1987 \JPA {\bf 20} 6623
\bibitem{Gra92} Grassberger P 1992 \JPA {\bf 25} 5867
\bibitem{Gau90} Gaunt D S and Colby S A 1990 {\it J.\ Stat.\ Phys.} {\bf 58} 539
\bibitem{Wan90a} Wang Z G, Nemirovsky A M, Freed K F and Myers K R 1990
\JPA {\bf 23} 2575
\bibitem{Sax87b} Saxena V K 1987 \PR B {\bf 35} 3612
\bibitem{Lar86} Larsson T A 1986 \JPA {\bf 19} 1691
\bibitem{Mon89} Mon K K and Vall\'{e}s J L 1989 \PR B {\bf 40} 2419
\bibitem{Lai89a} Lai P Y and Mon K K 1989 \JPA {\bf 22} 5167
\bibitem{Pri88} Privman V 1988 \PR B {\bf 38} 9261
\bibitem{Lai89b} Lai P Y and Mon K K 1989 \PR B {\bf 39} 12407
\bibitem{Car88} Cardy J L and Peschel I 1988 \NP {\bf B300 [FS22]} 377
\bibitem{Kac01} Kachkachi H and Garanin D A 2001 {\it Physica} A {\bf 300} 487 
\bibitem{Ple98b} Pleimling M and Selke W 1998 {\it Eur.\ Phys.\ J.} B {\bf 5} 805
\bibitem{Ple02a} Pleimling M 2002 {\it Comp.\ Phys.\ Commun.} {\bf 147} 101
\bibitem{Ple99} Pleimling M and Selke W 1999 \PR B {\bf 59} 65
\bibitem{Ple00} Pleimling M and Selke W 2000 \PR E {\bf 61} 933
\bibitem{Bro04} Brovchenko I, Geiger A and Oleinikova A 2004 {\it cond-mat/0402030}
\bibitem{Fis67} Fisher M E and Ferdinand F E  1967 \PRL {\bf 19} 169 
\bibitem{Bar79} Bariev R Z 1979 {\it Sov.\ Phys.\ JETP} {\bf 50} 613 
\bibitem{McC80} McCoy B M and Perk J H H 1980 \PRL {\bf 44} 840
\bibitem{Ko85} Ko L F, Au-Yang H and Perk J H H 1985 \PRL {\bf 54} 1091
\bibitem{Tur85} Turban L 1985 \JPA {\bf 18} L325
\bibitem{Del94} Delfino G, Mussardo G and Simonetti P 1994 \NP {\bf B432 [FS]} 518
\bibitem{Osh96} Oshikawa M and Affleck I 1996 \PRL {\bf 77} 2604
\bibitem{Sim98} Sim\~{o}es C S and Drugowich de Fel\'{i}cio J R 1998 \JPA {\bf 31} 7265
\bibitem{Sza99} Szalma F and Igl\'{o}i F 1999 {\it J. Stat. Phys.} {\bf 95} 795
\bibitem{Hen88b} Henkel M and Patk\'{o}s A 1988 \JPA {\bf 21} L231  
\bibitem{Hen89} Henkel M, Patk\'{o}s A and Schlottmann M 1989 \NP {\bf B314} 609
\bibitem{Tur04} Turban L {\it private communication}
\bibitem{Cos03} Costa-Santos R 2003 \PR B {\bf 68} 224423
\bibitem{Fro98} Fr\"{o}jdh P, Howard M and Lauritsen K B 1998 \JPA {\bf 31} 2311
\bibitem{Jan88} Janssen H K, Schaub B and Schmittmann B 1988 {\it Z.\ Phys.} B {\bf 72} 111
\bibitem{Kai94a} Kaiser C and Turban L 1994 \JPA {\bf 27} L579
\bibitem{Kai95} Kaiser C and Turban L 1995 \JPA {\bf 28} 351
\bibitem{Lau98} Lauritsen K B, Fr\"{o}jdh P and Howard M 1998 \PRL {\bf 81} 2104 
\bibitem{How00} Howard M, Fr\"{o}jdh P and Lauritsen K B 2000 \PR E {\bf 61} 167 
\bibitem{Han99} Hanke A, Krech M, Schlesener F and Dietrich S 1999 \PR E {\bf 60} 5163
\bibitem{Hau92} Hauge E H 1992 \PR A {\bf 46} 4994
\bibitem{Nap92} Napi\'{o}rkowski M, Koch W and Dietrich S 1992 \PR A {\bf 45} 5760 
\bibitem{Rej99} Rejmer K, Dietrich S and Napi\'{o}rkowski M 1999 \PR E {\bf 60} 4027
\bibitem{Par99} Parry A O, Rasc\'{o}n C and Wood A J 1999 \PRL {\bf 83} 5535
\bibitem{Par00a} Parry A O, Rasc\'{o}n C and Wood A J 2000 \PRL {\bf 85} 345
\bibitem{Par00b} Parry A O, Wood A J and Rasc\'{o}n C 2000 \JPCM {\bf 12} 7671
\bibitem{Par01a} Parry A O, Wood A J, Carlon E and Drzewi\'{n}ski A 2001 \PRL {\bf 87} 196103
\bibitem{Par01b} Parry A O, Wood A J and Rasc\'{o}n C 2001 \JPCM {\bf 13} 4591
\bibitem{Bed01} Bednorz A and Napi\'{o}rkowski M 2001 \PR E {\bf 63} 031602
\bibitem{Lip98} Lipowski A 1998 \PR E {\bf 58}, R1 
\bibitem{Abr02} Abraham D B and Macio{l}ek A 2002 \PRL {\bf 89} 286101
\bibitem{Alb03} Albano E V, De Virgiliis A, M\"{u}ller M and Binder K 2003 \JPCM {\bf 15} 333
\bibitem{Mil03} Milchev A, M\"{u}ller M, Binder K and Landau D P 2003 \PRL {\bf 90} 136101 
\bibitem{Die90a} Diehl H W and N\"{u}sser A 1990 {\it Z.\ Phys.} B {\bf 79} 69
\bibitem{Ben85} Benyoussef A, Boccara N and Saber M 1985 \JPC {\it 18} 4275
\bibitem{Sab87} Saber M 1987 \JPC {\bf 20} 2749
\bibitem{Dak01} Dakhama A, Fahti A and Benayad N 2001 {\it Eur.\ Phys.\ J.} B {\bf 21} 393
\bibitem{Kan88} Kaneyoshi T 1988 {\it Phys.\ Stat.\ Sol.} (b) {\bf 150} 297
\bibitem{Kan89} Kaneyoshi T 1989 \PR B {\bf 39} 557 
\bibitem{Kan91} Kaneyoshi T 1991 {\sl Introduction to surface magnetism},
(Boca Raton, Ann Arbor, and Boston: CRC Press) 
\bibitem{Wan90b} Wang J-S, Selke W, Dotsenko Vl S and
Andreichenko V B 1990 {\it Physica} A {\bf 164} 221
\bibitem{Kan98} Kaneyoshi T and Shin S 1998 {\it Physica} A {\bf 260} 455
\bibitem{Mon88} Mon K K and Nightingale M P 1988 \PR B {\bf 37} 3815
\bibitem{Dur87} Durian D J and Franck C 1987 \PRL {\bf 59} 555
\bibitem{Har74a} Harris A B 1974 \JPC {\bf 7} 1671
\bibitem{Das88} Da Silva L R , Tsallis C and Sarmento E F 1988 \PR B {\bf 37} 7832
\bibitem{Sha94} Shalaev B N 1994 {\it Phys.\ Rep.} {\bf 237} 129
\bibitem{Fel02} Feldman D E and Vinokur V M 2002 \PRL {\bf 89} 227204
\bibitem{Igl91} Igl\'{o}i F, Turban L and Berche B 1991 \JPA {\bf 24} L1031
\bibitem{Car91} Cardy J 1991 \JPA {\bf 24} L1315
\bibitem{DeM98} De Martino A, Moriconi M and Mussardo G 1998 \NP {\bf B509 [FS]} 615
\bibitem{Bag04} Pleimling M, Bagam\'{e}ry F \'{A}, Turban L and Igl\'{o}i F 2004 {\it cond-mat/0402188}
\bibitem{Nat88} Nattermann T and Villain J 1988 {\it Phase Transitions} {\bf 11} 5
\bibitem{Eis82} Eisenriegler E, Kremer K and Binder K 1982 {\it J.\ Chem.\ Phys.} {\bf 77} 6296
\bibitem{Die90b} Diehl H W and N\"{u}sser A 1990 {\it Z.\ Phys.} B {\bf 79} 79
\bibitem{Die98b} Diehl H W 1998 {\it Eur.\ Phys.\ J.} B {\bf 1} 401
\bibitem{Ben99} Bengrine M, Benyoussef A, Ez-Zahraouy H,
and Mhirech F 1999 {\it Physica} A {\bf 268} 149
\bibitem{Zha00} Zhao D, Feng L, Huber D L  and Lagally M G 2000
\PR B {\bf 62} 11316
\bibitem{Mok01} Mokrani A and Vega A 2001 \PR B {\bf 63} 094403 
\bibitem{Bah02a} Bahmad L, Benyoussef A and Ez-Zahraouy H 2002 {\it Physica} {\bf 303} 525
\bibitem{Rei98a} Aar\~{a}o Reis F D A 1998 \PR B {\bf 58} 394
\bibitem{Rei00} Aar\~{a}o Reis F D A 2000 \PR B {\bf 62} 6565
\bibitem{Han01} Hanke A and Kardar M 2001 \PRL {\bf 86} 4596
\bibitem{Han02} Hanke A and Kardar M 2002 \PR E {\bf 65} 046121
\bibitem{Dup98} Duplantier B 1998 \PRL {\bf 81} 5489
\bibitem{Car99} Cardy J 1999 \JPA {\bf 32} L177
\bibitem{Hil81} Hilhorst H J and van Leeuwen J M  1981 \PRL {\bf 47} 1188
\bibitem{Har74b} Harris A B and Lubensky T C 1974 \PRL {\bf 33} 1540
\bibitem{Gri76} Grinstein G and Luther A 1976 \PR B {\bf 13} 1329
\bibitem{Ohn92} Ohna K and Okabe Y 1992 \PR B {\bf 46} 5917
\bibitem{Usa01} Usatenko Z, Shpot M and Hu C-K 2001 \PR E {\bf 63} 056102
\bibitem{Usa02} Usatenko Z and Hu C-K 2002 \PR E {\bf 65} 066103
\bibitem{Usa03} Usatenko Z and Hu C-K 2003 \PR E {\bf 68} 066115
\bibitem{Igl98} Igl\'{o}i F, Lajk\'{o} P, Selke W and Szalma F 1998
\JPA {\bf 31} 2801 
\bibitem{Laj00} Lajk\'{o} P and Igl\'{o}i F 2000 \PR E {\bf 61} 147
\bibitem{Pal00} Pal\'{a}gyi G, Chatelain C, Berche B and F.\ Igl\'{o}i F
2000 {\it Eur.\ Phys.\ J.} B {\bf 13} 357
\bibitem{Rei97} Aar\~{a}o Reis F D A 1997 \PR B {\bf 55} 11084
\bibitem{Rei98b} Aar\~{a}o Reis F D A 1998 {\it Physica} A {\bf 257} 495
\bibitem{Rei98c} Aar\~{a}o Reis F D A 1998 \JPA {\bf 31} 9105
\bibitem{Rei01} Aar\~{a}o Reis F D A 2001 {\it Physica} A {\bf 291} 375
\bibitem{She97} Shen J, Skomski R, Klaua M, Jenniches H, 
Manoharan S S and Kirschner J 1997 \PR B {\bf 56} 2340
\bibitem{She03} Shen J, Pierce J P, Plummer E W and Kirschner J 2003 \JPCM {\bf 15} R1
\bibitem{Sko03} Skomski R 2003 \JPCM {\bf 15} R841
\bibitem{Baz00} Bazhanov D I, Hergert W, Stepanyuk V S, Katsnelson A A,
Rennert P, Kokko K and Demangeat C 2000 \PR B {\bf 62} 6415
\bibitem{Bel01} Bellini V, Papanikolaou N, Zeller R and Dederichs P H
2001 \PR B {\bf 64} 094403
\bibitem{And01} Andrieu S, Chatelain C, Lemine M, Berche B and
Bauer P 2001 \PRL {\bf 86} 3883
\bibitem{Bar90} Bariev R Z, Malov O A and Barieva N A 1990 {\it Physica} A {\bf 169} 281
\bibitem{Ber91} Berche B and Turban L 1991 \JPA {\bf 24} 245
\bibitem{Igl90} Igl\'{o}i F and Indekeu J O 1990 \PR B {\bf 41} 6836
\bibitem{Ham03} Hamedoun M, Bakrim H, Hourmatallah A and Benzakour N 2003
{\it Surface Science} {\bf 539} 155
\bibitem{Zha02} Zhao D, Liu F, Huber D L and Lagally M G 2002 
{\it J.\ Appl.\ Phys.} {\bf 91} 3150
\end{thebibliography}
\end{document}